\documentclass[11pt,ams]{article}%
\usepackage{geometry}
\usepackage{amsmath}
\usepackage{amsfonts}
\usepackage{amssymb}
\usepackage{graphicx}%

\begin{document}

\date{}
\title{\textbf{Gluon confinement, $i$-particles and BRST soft breaking}}
\author{\textbf{S.\,P.\,\,Sorella}\thanks{sorella@uerj.br}\ \thanks{Work supported by
FAPERJ, Funda{\c{c}}{\~{a}}o de Amparo {\`{a}} Pesquisa do Estado do Rio de
Janeiro, under the program \textit{Cientista do Nosso Estado},
E-26/101.578/2010.}\\[0.5cm]
\textit{{\small {UERJ $-$ Universidade do Estado do Rio de
Janeiro}}}\\\textit{{\small {Instituto de F\'{\i}sica $-$
Departamento de F\'{\i}sica Te\'{o}rica}}}\\\textit{{\small {Rua
S{\~a}o Francisco Xavier 524, 20550-013 Maracan{\~a}, Rio de
Janeiro, Brasil}}}$$} \maketitle

\begin{abstract}
\noindent A few issues on gluon confinement are addressed with the help of a renormalizable gauge model obtained by introducing a replica of the Faddeev-Popov action and a soft breaking of the BRST symmetry. Confinement turns out to be encoded in the spectral properties of the corresponding correlation functions. While the propagators of the elementary fields have no interpretation in terms of physical excitations, examples of local composite operators whose two-point correlation functions possess a spectral representation with positive spectral density can be introduced. These composite operators turn out to be left invariant by the BRST transformations, a feature which has strong consequences on their renormalizability properties. Moreover, they display a direct interpretation in terms of $i$-particles \cite{Baulieu:2009ha}, which are the unphysical modes corresponding to a confining propagator of the Gribov type. A possible way to take into account the effects of the Gribov copies is also outlined.
\end{abstract}

\baselineskip=13pt

\newpage

\section{Introduction}
Color confinement is a fascinating topic, representing a great challenge from theoretical, numerical and experimental points of view. Although impressive results have been achieved in the last decades, it is safe to state that many aspects remain still to be unraveled, forcing us to provide new ideas and to improve our current computational capabilities.\\\\The aim of the present work is twofold. Our first goal is that of investigating confinement within the framework of a local and renormalizable Euclidean quantum field theory. Here, confinement is meant to be encoded in the analyticity properties of the correlation functions, as expressed by the K\"all\'{e}n-Lehmann spectral representation, which provides a powerful tool in order to attach a physical meaning to the correlation functions \cite{Itzykson:1980rh,Weinberg:1995mt}. More precisely, in a confining theory, we expect that the two-point correlation functions of the elementary fields, here generically denoted by $\{\phi\}$, cannot be cast in the form of a spectral representation with positive spectral density. A meaningful particle interpretation of the propagators of the theory cannot thus be achieved. As a consequence, the excitations corresponding to the elementary fields $\{\phi\}$ do not belong to the physical spectrum of the theory. As a familiar example of this, let us quote the Gribov propagator \cite{Gribov:1977wm}, which we shall employ extensively throughout:
\begin{equation}
\langle \phi(k) \phi(-k) \rangle = \frac{k^2}{k^4+\gamma^4} \;, \label{gbp}
\end{equation}
where $k^2\ge 0$ is the Euclidean momentum and $\gamma$ is the so called Gribov mass parameter. This kind of propagator arises in the Gribov-Zwanziger theory \cite{Gribov:1977wm,Zwanziger:1988jt,Zwanziger:1989mf,Zwanziger:1992qr},  which implements in a renormalizable way the restriction of the domain of integration in the Euclidean Faddeev-Popov quantization formula to the interior of the region in field space bounded by the Gribov horizon\footnote{See ref.\cite{Dudal:2009bf} for a general account on the Gribov-Zwanziger theory.}.  Expression \eqref{gbp} displays complex poles, so that it cannot be interpreted as the propagator of a physical particle, being suitable for a confining phase. It can be seen as describing the propagation of two unphysical modes with imaginary masses $\pm i{ \gamma}^2$, which have been called $i$-particles in \cite{Baulieu:2009ha}, namely
\begin{equation}
\frac{k^2}{k^4+\gamma^4} = \frac{1}{2} \left( \frac{1}{k^2-i\gamma^2} + \frac{1}{k^2+i\gamma^2} \right) \;. \label{ip}
\end{equation}
Furthermore, a suitable set of local composite operators, $\{ O(\phi) \}$, should  be introduced in such a way that the two-point correlation functions, $\langle O(k) O(-k) \rangle$, exhibit the K\"all\'{e}n-Lehmann spectral  representation
\begin{equation}
\langle O(k) O(-k) \rangle = \int_{\tau_0}^{\infty} d\tau \; \frac{\rho(\tau)}{\tau + k^2}  \;, \label{klr}
\end{equation}
where $\rho(\tau)$ is the positive spectral density and $\tau_0 \ge 0$ stands for the threshold. As it follows from complex analysis, expression \eqref{klr} allows us to define an analytic function in the complex cut $k^2$-plane \cite{Itzykson:1980rh,Weinberg:1995mt}, where the cut extends along the negative real axis, from $-\tau_0$ to $-\infty$.   An important outcome of expression \eqref{klr} is represented by the welcome feature that we can move from Euclidean to Minkowski space, {\it i.e.} $k^2 \rightarrow -k^2_M$, where the cut will be now located along the positive real axis, starting at the threshold $+\tau_0$ and extending till $+\infty$. Moreover, positivity of the spectral density $\rho(\tau)$ enables us to give an interpretation of eq.\eqref{klr} in terms of physical states with positive norm\footnote{The ideal situation would be that in which, in addition to cuts along the negative real axis, we would  be able to isolate a real pole in the correlation function $\langle O(k) O(-k) \rangle$, {\it i.e.} 
\begin{equation}
\langle O(k) O(-k) \rangle = \frac{{\cal R}(k^2)}{k^2+M^2} + \int_{\tau_0}^{\infty} d\tau \; \frac{\rho(\tau)}{\tau + k^2}  \;, \label{klr1}
\end{equation}
where $M$ is a mass parameter and ${\cal R}(k^2)$ an analytic function. Such an expression would provide a direct link with the physical spectrum of the theory. The mass $M$ would refer to the mass of a physical excitation created by the action of the composite operator $O(\phi)$ on the vacuum of the theory. This is what one expects for the spectrum of the glueballs in $QCD$, which should manifest as poles in the correlation functions of suitable gauge invariant composite operators \cite{Teper:1998kw}. Nevertheless, from an analytic point of view, finding a pole is a highly non-trivial matter. To the best of our knowledge, this would require a re-summation of an infinite set of diagrams \cite{Itzykson:1980rh,Weinberg:1995mt}, a not easy task indeed. A more modest plan, which will be assumed here, would be that of not facing the too hard  calculation of the pole, limiting ourselves to the determination of the cuts along the negative real axis,  something which can be already regarded as a non-trivial achievement, given the complexity of the Gribov propagator \eqref{gbp}. Though, we might also argue that the knowledge of the threshold $\tau_0$ could provide us a rough estimate of the mass $M$, as follows by observing that $\tau_0$ should be related to the energy for the creation of  multi-particle states \cite{Itzykson:1980rh,Weinberg:1995mt}. }. The operators $\{ O(\phi) \}$ would thus provide information on the spectrum of the theory.  \\\\One should notice that the correlation function $\langle O(k) O(-k) \rangle$ is supposed to be evaluated by employing a confining propagator of the Gribov type, eq.\eqref{gbp}, which exhibits complex poles. It is thus not obvious that expression \eqref{klr} will display the desired analyticity properties. \\\\To have an idea of the difficulty related to the use of a confining propagator of the Gribov type, let us quote the case of the two-point correlation function $\langle F^2(x) F^2(y) \rangle$, $F^2(x)=F^a_{\mu\nu}(x) F^a_{\mu\nu}(x)$, which has been investigated at one-loop order in \cite{Zwanziger:1989mf,Gracey:2009mj}. In spite of the fact that  $F^2(x)$ is a BRST invariant and renormalizable operator within the Gribov-Zwanziger theory \cite{Dudal:2009zh}, it gives rise to a correlation function which, besides  exhibiting a cut along the negative real axis, does have unphysical cuts along the imaginary axis \cite{Zwanziger:1989mf}. A promising way out in order to avoid the presence of the unphysical cuts has been outlined in \cite{Baulieu:2009ha}, where  the introduction of the $i$-particles has been proven to be  helpful in order to construct examples of composite operators whose correlation functions exhibit the K\"all\'{e}n-Lehmann spectral representation at one-loop order. Though, it turns out that, unlike $F^2(x)$, these composite operators are not left invariant by the softly broken BRST symmetry which ensures the renormalizability of the Gribov-Zwanziger action \cite{Dudal:2009bf}. \\\\Further work seems to be needed in order to achieve a deeper understanding of the relationship between the softly broken BRST symmetry and the construction of a sensible set of renormalizable composite operators exhibiting good analyticity properties. This is  the second topic we aim to investigate  in  the present paper. We shall elaborate on a model which, in analogy with the Gribov-Zwanziger theory, displays a softly broken BRST symmetry. Moreover, we shall be able to introduce examples of composite operators  which are invariant under the softly broken BRST symmetry, while having a direct interpretation in terms of $i$-particles. This feature enables us to immediately establish the K\"all\'{e}n-Lehmann spectral representation for the corresponding correlation functions at one-loop order. In addition, the invariance property of these operators translates into softly broken Slavnov-Taylor identities which play a central role in order to establish how these operators renormalize, {\it i.e.} how they mix under radiative corrections. The knowledge of the mixing matrix is needed in order to correctly evaluate the correlation functions at higher orders as well as to find out quantum operators which are invariant under the renormalization group equations, see \cite{Dudal:2008tg,Dudal:2009zh} and references therein. \\\\We  point out that the issue of the existence of a BRST symmetry compatible with the confining character of the theory is one of the highlight of the current investigation on the infrared behavior of Yang-Mills theories and, in particular, on the possible existence of a non-perturbative version of the BRST symmetry. We remind here that this symmetry plays a fundamental role in the proof of the renormalizability and of the unitarity of non-Abelian gauge theories in the perturbative regime. At present, it is unknown if and how such a powerful symmetry can be reconciled with the existence of the Gribov copies.  What emerges from the studies of the Gribov-Zwanziger theory  is that the  BRST symmetry enjoyed by the Faddeev-Popov action is softly broken by  terms  which are proportional to the Gribov parameter $\gamma$ \cite{Dudal:2008sp,Baulieu:2008fy,Baulieu:2009xr}. As such, the breaking of the BRST symmetry seems to be unavoidably linked to the presence of the Gribov horizon. Despite the various efforts done recently \cite{Fischer:2008uz,Sorella:2009vt,Kondo:2009qz,Ilderton:2009qc}, the issue of the BRST versus the Gribov horizon is certainly far from being understood. Therefore, it seems to us that the construction of a confining  model exhibiting a softly broken BRST symmetry and allowing for the introduction of examples of composite operators which are BRST invariant  and whose correlation functions display the K\"all\'{e}n-Lehmann spectral representation might be of some help for the current debate. \\\\The paper is organized as follows. In Sect.\ref{gm} we present the model and its softly broken BRST symmetry. We motivate its introduction by showing that it has a natural interpretation in terms of $i$-particles \cite{Baulieu:2009ha}, {\it i.e.} in terms of the unphysical modes corresponding to the confining Gribov propagator, eq.\eqref{gbp}. In Sect.\ref{decent} we provide an example of a local composite operator whose two-point correlation function exhibits the  K\"all\'{e}n-Lehmann representation at one-loop order. This composite operator turns out to be left invariant by the softly broken BRST symmetry, a feature which is relevant for the study of its renormalizability, which we shall briefly outline. Sect.\ref{Gribov} deals with the issue of the Gribov copies. A possible set up, relying on the use of Gribov's no-pole condition \cite{Gribov:1977wm}, is analysed in order to account for the Gribov phenomenon. In Sect.\ref{refined} we present a refined version of the model which, in analogy to the so-called refined Gribov-Zwanziger action introduced in \cite{Dudal:2007cw,Dudal:2008sp}, gives rise to a modified gluon propagator attaining a finite non-vanishing value at zero momentum,  a behavior also reported by the analysis of the Schwinger-Dyson equations  \cite{Aguilar:2004sw,Aguilar:2008xm,Boucaud:2008ky,Fischer:2008uz} as well as by models employing an effective gluon mass  \cite{Cornwall:1981zr,Tissier:2010ts}. We mention here that  recent lattice numerical simulations  \cite{Cucchieri:2007rg,Cucchieri:2008fc,Cucchieri:2008mv,Cucchieri:2009zt,Bogolubsky:2009dc,Bogolubsky:2009qb,Dudal:2010tf} point towards a gluon propagator exhibiting a finite non-vanishing value at zero momentum. Sect.\ref{conclusion} collects our conclusion. A few technical details about the Ward identities fulfilled by the model are given in the Appendix \ref{wid}.

\section{The model and its softly broken BRST symmetry}  \label{gm}

The construction of the gauge model which we shall present in the following relies on the mastering of the confining Gribov type propagator, eq.\eqref{gbp}, which we have been able to acquire so far \cite{Baulieu:2009ha}.

\subsection{A confining gauge model}

Let  us start by considering the Faddeev-Popov action in the Landau gauge
\begin{equation}
S_{FP} = \int d^4x\; \left(  \frac{1}{4} F^a_{\mu\nu}(A) F^a_{\mu\nu}(A) + b^a \partial_\mu A^a_\mu +
{\bar c}^a \partial_{\mu} D^{ab}_\mu(A) c^b \right)  \;, \label{fpact}
\end{equation}
where $b^a$ is the Lagrange multiplier enforcing the Landau gauge condition, $\partial_{\mu} A^a_{\mu}=0$, $({\bar c}^a, c^a)$ stand for the Faddeev-Popov ghosts, $F^a_{\mu\nu}(A)$ is the field strength
\begin{equation}
F^a_{\mu\nu}(A) = \partial_{\mu} A^a_{\nu} - \partial_{\nu} A^a_{\mu} + g f^{abc} A^b_{\mu} A^c_{\nu} \;, \label{fst}
\end{equation}
and $D^{ab}_\mu(A) $ denotes the covariant derivative
\begin{equation}
D^{ab}_\mu(A) c^b = \partial_\mu c^a + g f^{acb} A^c_\mu c^b \;. \label{cda}
\end{equation}
Our confining model is obtained by performing the following steps:
\begin{itemize}
\item we first consider a replica of the Faddeev-Popov action, eq.\eqref{fpact}, by introducing a set of mirror fields $(U^a_\mu, {\bar b}^a, {\bar \omega}^a, \omega^a)$ as well as a mirror Faddeev-Popov action:
\begin{equation}
S_{MFP} = \int d^4x\; \left(  \frac{1}{4} U^a_{\mu\nu}(U) U^a_{\mu\nu}(U) + {\bar b}^a \partial_\mu U^a_\mu +
{\bar \omega}^a \partial_{\mu} D^{ab}_\mu(U) \omega^b \right)  \;, \label{mfpact}
\end{equation}
where $U^a_{\mu\nu}(U)$ is the field strength corresponding to the field $U^a_\mu$
\begin{equation}
U^a_{\mu\nu}(U) = \partial_{\mu} U^a_{\nu} - \partial_{\nu} U^a_{\mu} + g f^{abc} U^b_{\mu} U^c_{\nu} \;, \label{mfst}
\end{equation}
while ${\bar b}^a$ and $({\bar \omega}^a, \omega^a)$ stand for the mirror Lagrange multiplier and mirror Faddeev-Popov ghosts, respectively, and
\begin{equation}
D^{ab}_\mu(U) \omega^b = \partial_\mu \omega^a + g f^{acb} U^c_\mu \omega^b \;. \label{mcda}
\end{equation}
\item the gauge field $A^a_\mu$ is softly coupled to its mirror field $U^a_\mu$ through the following mixed term
\begin{equation}
S_{\theta} = i \sqrt{2} \theta^2 \int d^4x\; A^a_\mu U^a_\mu \;, \label{mass}
\end{equation}
where $\theta$ is a mass parameter which, for the time being, is considered as a free parameter introduced by hand. As we shall see, this mass parameter will play a role akin to that of the Gribov parameter $\gamma$ of the Gribov-Zwanziger action \cite{Gribov:1977wm,Zwanziger:1988jt,Zwanziger:1989mf,Zwanziger:1992qr}. Later on, in Sect.\ref{Gribov}, we shall attempt to provide a possible dynamical mechanism for $\theta$.
\end{itemize}
Our confining model is thus specified by the following action
\begin{eqnarray}
S & = & S_{FP} + S_{MFP} + S_{\theta}  \nonumber \\
    & = &  \int d^4x\; \left(  \frac{1}{4} F^a_{\mu\nu}(A) F^a_{\mu\nu}(A) + \frac{1}{4} U^a_{\mu\nu}(U) U^a_{\mu\nu}(U)
     + i \sqrt{2} \theta^2 A^a_\mu U^a_\mu \right) \nonumber \\
    & + &  \int d^4x \; \left(b^a \partial_\mu A^a_\mu + {\bar c}^a \partial_{\mu} D^{ab}_\mu(A) c^b
    +  {\bar b}^a \partial_\mu U^a_\mu + {\bar \omega}^a \partial_{\mu} D^{ab}_\mu(U) \omega^b \right)  \;. \label{model}
\end{eqnarray}
A first property of the action \eqref{model} follows by looking at the propagators of the fields $(A^a_{\mu}, U^a_\mu)$, {\it i.e} 
\begin{equation}
\langle A^a_\mu(k) A^b_\nu(-k) \rangle = \delta^{ab} \frac{k^2}{k^4+2\theta^4} \left( \delta_{\mu\nu} -\frac{k_\mu k_\nu}{k^2} \right) \;, \label{paa}
\end{equation}
\begin{equation}
\langle U^a_\mu(k) U^b_\nu(-k) \rangle = \delta^{ab} \frac{k^2}{k^4+2\theta^4} \left( \delta_{\mu\nu} -\frac{k_\mu k_\nu}{k^2} \right) \;, \label{puu}
\end{equation}
\begin{equation}
\langle A^a_\mu(k) U^b_\nu(-k) \rangle = \delta^{ab} \frac{-i\sqrt{2}\theta^2}{k^4+2\theta^4} \left( \delta_{\mu\nu} -\frac{k_\mu k_\nu}{k^2} \right) \;. \label{pau}
\end{equation}
As one sees from expressions \eqref{paa}, \eqref{puu}, \eqref{pau}, all propagators are of the confining Gribov type. As such, they correspond to unphysical excitations. Said otherwise, we cannot attach a particle interpretation to the propagators of the elementary fields $(A^a_{\mu}, U^a_\mu)$. \\\\The second feature displayed by the model is that it contains a unique coupling constant $g$. Both fields $A^a_{\mu}$ and $U^a_\mu$ interact  with the same coupling. We shall provide further comments on this aspect when the relationship with the $i$-particles will be discussed. Nevertheless, we point out that the feature of having a unique coupling constant is protected by a powerful discrete invariance which we shall refer to as the mirror symmetry. It turns out in fact that the action $S$, eq.\eqref{model}, is left invariant by the following discrete transformations
\begin{eqnarray}
A^a_{\mu}  \rightarrow  U^a_{\mu} \;, \nonumber \\
U^a_{\mu}  \rightarrow A^a_{\mu} \;, \nonumber \\
b^a  \rightarrow {\bar b}^a \;, \nonumber \\
{\bar b}^a  \rightarrow { b}^a \;, \nonumber \\
c^a  \rightarrow {\omega}^a \;, \nonumber \\
\omega^a  \rightarrow {c}^a \;, \nonumber \\
{\bar c}^a  \rightarrow {\bar \omega}^a \;, \nonumber \\
{\bar \omega}^a  \rightarrow {\bar c }^a \;. \label{mirror}
\end{eqnarray}
The mirror symmetry \eqref{mirror} means that the fields $(A^a_\mu, {b}^a, {\bar c}^a, c^a)$ can be replaced by  $(U^a_\mu, {\bar b}^a, {\bar \omega}^a, \omega^a)$, and vice-versa. Notice that the mirror symmetry would be lost if a second coupling, $g'$, would have been introduced. Suppose in fact that the field strength of the field $U^a_\mu$ would contain a different coupling, {\it i.e.}
\begin{equation}
U^a_{\mu\nu}(U) = \partial_{\mu} U^a_{\nu} - \partial_{\nu} U^a_{\mu} + g'  f^{abc} U^b_{\mu} U^c_{\nu} \;. \label{diffc}
\end{equation}
Evidently, the mirror symmetry would not hold anymore, as $U^a_{\mu\nu}(U)$ cannot be transformed into the mirror field strength  $F^a_{\mu\nu}(A)$.\\\\A third feature displayed by the action \eqref{model} is the existence of a softly broken BRST symmetry. It is easy to verify that the nilpotent BRST transformations
\begin{eqnarray}
s A^a_\mu & = & - D^{ab}_\mu(A) c^b \;, \nonumber \\
s U^a_\mu & = & - D^{ab}_\mu(U) \omega^b \;, \nonumber \\
s c^a & = & \frac{g}{2} f^{abc} c^b c^c \;, \nonumber \\
s \omega^a & = & \frac{g}{2} f^{abc} \omega^b \omega^c \;, \nonumber \\
s {\bar c}^a & = & b^a  \;, \nonumber \\
s b^a & = & 0 \;, \nonumber \\
s {\bar \omega}^a & = & {\bar b}^a  \;, \nonumber \\
s {\bar b}^a & = & 0 \;, \label{brst}
\end{eqnarray}
leave the action $S$ invariant up to soft terms proportional to the parameter $\theta^2$, {\it i.e.}
\begin{equation}
s S = \theta^2 \Delta_{break} \;, \label{break}
\end{equation}
where $\Delta_{break}$ is given by
\begin{equation}
\Delta_{break} = -i \sqrt{2} \int d^4x \; \left( U^a_\mu D^{ab}_\mu(A)c^b + A^a_\mu D^{ab}_\mu(U)\omega^b \right) \;. \label{bd}
\end{equation}
Being of dimension two in the fields, $\Delta_{break}$ is a soft breaking.

\subsection{Recovering Yang-Mills theory in the deep perturbative ultraviolet region}

An important aspect of the Gribov-Zwanziger action is that the perturbative ultraviolet behavior of Yang-Mills theory is recovered when the Gribov horizon is removed. More precisely, when the Gribov parameter is set to zero, $\gamma=0$, the Gribov-Zwanziger action can be proven to be equivalent to the Faddeev-Popov action. This property of the Gribov-Zwanziger action is encoded in the BRST transformations of the auxiliary fields $(\phi^{ab}_{\mu}, {\bar \phi}^{ab}_\mu, \omega^{ab}_{\mu}, {\bar \omega}^{ab}_{\mu} )$ needed to implement the restriction to the Gribov region \cite{Dudal:2009bf}. These fields form a BRST quartet, thus yielding a trivial BRST cohomology. As the BRST symmetry becomes an exact symmetry when $\gamma=0$, it follows that the BRST quartet decouples from the theory, so that the Faddeev-Popov action is recovered at the end, see \cite{Dudal:2009bf}. \\\\This property is also displayed by our action $S$, eq.\eqref{model}, although the recovering mechanism is deeply different from that of the Gribov-Zwanziger theory. One notices  that, when $\theta=0$, the BRST transformations of the mirror fields  $(U^a_\mu, {\bar b}^a, {\bar \omega}^a, \omega^a)$, eqs.\eqref{brst}, remain non-trivial, {\it i.e.} they give rise to nonvanishing BRST cohomology classes \cite{Piguet:1995er}.  However, it turns out that, when $\theta=0$, the mirror fields $(U^a_\mu, {\bar b}^a, {\bar \omega}^a, \omega^a)$ do not interact with $(A^a_\mu, {b}^a, {\bar c}^a, c^a)$. In fact, by construction, when the mass parameter $\theta$ is set to zero, the two set of fields $(A^a_\mu, {b}^a, {\bar c}^a, c^a)$ and $(U^a_\mu, {\bar b}^a, {\bar \omega}^a, \omega^a)$ belong to two disjoint theories which do not interact, namely
\begin{equation}
S\Bigl|_{\theta=0}= S_{FP}(A,b,{\bar c},c) + S_{MFP}(U,{\bar b},{\bar \omega},\omega)  \;. \label{plm}
\end{equation}
As a consequence, denoting by $\left( O_1(A,b,{\bar c},c),....,O_N(A,b,{\bar c},c)\right)$ a set of operators depending only on the original fields $(A^a_\mu, {b}^a, {\bar c}^a, c^a)$, we have
\begin{eqnarray} 
\langle O_1(A,b,{\bar c},c)....O_N(A,b,{\bar c},c) \rangle \Bigl|_{\theta=0} & = & \frac{\int\; [{\cal D}\Phi] \; O_1(A,b,{\bar c},c)....O_N(A,b,{\bar c},c) \; e^{-\left( S_{FP}(A,b,{\bar c},c) + S_{MFP}(U,{\bar b},{\bar \omega},\omega) \right)}}{\int\; [{\cal D}\Phi] \; e^{-\left( S_{FP}(A,b,{\bar c},c) + S_{MFP}(U,{\bar b},{\bar \omega},\omega) \right)}} \nonumber \\
& = &  \frac{\int \; {\cal D}A {\cal D}b {\cal D}c {\cal D}{\bar c}\; O_1(A,b,{\bar c},c)....O_N(A,b,{\bar c},c) \; e^{- S_{FP}(A,b,{\bar c},c)  }}{\int {\cal D}A {\cal D}b {\cal D}c {\cal D}{\bar c} \; e^{- S_{FP}(A,b,{\bar c},c)  }} \;, \nonumber \\
 \label{equiv}
\end{eqnarray}
from which one sees that the Faddeev-Popov theory is recovered when $\theta=0$. The usual perturbation theory is thus re-obtained in the deep ultraviolet region, where the soft parameter $\theta$ becomes negligible.

\subsection{Renormalizability and Slavnov-Taylor identities} \label{ren}

\subsubsection{Properties of the Gribov propagator and non-renormalization of the soft mass term} 

Before showing how the softly broken BRST transformations, eqs.(\ref{brst}), can be employed to derive suitable Slavnov-Taylor identities, let us give a simple argument in order to convince the reader of the renormalizability of expression  \eqref{model}. The reasoning is based on the use of the dimensional regularization\footnote{Notice that, in the present case, dimensional regularization in an invariant regularization.} with minimal subtraction and on the following property of the Gribov propagator
\begin{equation} 
\frac{k^2}{k^4+2\theta^4} = \frac{1}{k^2} -\frac{2\theta^4}{k^2(k^4+2\theta^4)} \;. \label{p1gb}
\end{equation} 
Replacing each Gribov propagator in the one-loop diagrams by expression \eqref{p1gb} enables us to infer that no divergent terms proportional to the soft parameter $\theta^2$ might arise. Let us discuss explicitly the absence of one-loop counterterms of the form $\theta^2 A^a_\mu A^a_\mu$ and $\theta^2 U^a_\mu U^a_\mu$. These counterterms would arise, for example,  from tadpole diagrams in the two point $A$-$A$  and $U$-$U$ $1PI$ Green's functions, and would be related to the momentum integral 
\begin{equation} 
\int \frac{d^dk}{(2\pi)^d} \frac{k^2}{k^4+2\theta^4} \;, \label{td}
\end{equation} 
 where $d=4-\varepsilon$. From identity \eqref{p1gb}, it follows 
 \begin{equation} 
 \int \frac{d^dk}{(2\pi)^d} \frac{k^2}{k^4+2\theta^4} =  \int \frac{d^dk}{(2\pi)^d} \frac{1}{k^2}-  \int \frac{d^dk}{(2\pi)^d} \frac{2\theta^4}{k^2(k^4+2\theta^4)}  
  =  - 2\theta^4  \int \frac{d^dk}{(2\pi)^d} \frac{1}{k^2(k^4+2\theta^4)}  \;, \label{fr}
 \end{equation}
where use has been made of dimensional regularization. The right-hand side of equation \eqref{fr} is convergent in the ultraviolet region by power counting, showing thus that no divergent terms proportional to $\theta^2$ arise\footnote{It is worth pointing out  that the result in eq.\eqref{fr} is deeply related to the form of the Gribov propagator. Repeating the same argument for a Yukawa type propagator would give rise to a mass dependent divergent term. In fact
\begin{equation} 
 \int \frac{d^dk}{(2\pi)^d} \frac{1}{k^2+\lambda^2} =  \int \frac{d^dk}{(2\pi)^d} \frac{1}{k^2}-  \int \frac{d^dk}{(2\pi)^d} \frac{\lambda^2}{k^2(k^2+\lambda^2)}  
  =  - \lambda^2 \int \frac{d^dk}{(2\pi)^d} \frac{1}{k^2(k^2+\lambda^2)}  \;, \label{fr1}
 \end{equation}
which is divergent in the ultraviolet region for $d=4-\varepsilon$.}. Similarly, one can state that no counterterm of the type $\theta^2 A^a_\mu U^a_\mu$ is required for the one-loop two point mixed $A$-$U$ Green's function, as it follows by noticing that the corresponding Feynman integral is ultraviolet convergent by power counting. \\\\Since no counterterms proportional to $\theta^2$ can arise at one-loop order, it follows that all possible divergences occurring in the model are those corresponding to $\theta^2=0$, in which case the action $S$ reduces to that of two completely disjoint expressions, see eq.\eqref{plm}. As a consequence, all possible divergences are those of the familiar Yang-Mills theory in the Landau gauge, see  \cite{Dudal:2008sp,Dudal:2009bf} and references therein. Moreover, taking into account the mirror symmetry, it follows that the renormalization factors of the mirror fields $(U^a_\mu, {\bar b}^a, {\bar \omega}^a, \omega^a)$ are the same as those of $(A^a_\mu, {b}^a, {\bar c}^a, c^a)$ which, in the Landau gauge, are given by \cite{Dudal:2008sp,Dudal:2009bf}: 
\begin{eqnarray} 
Z_U & = & Z_A  \;, \nonumber \\
Z_{\bar b} & = & Z_b = Z_A^{-1} \;, \nonumber \\
Z_{\bar \omega} & = & Z_\omega= Z_{\bar c} = Z_c \;, \label{zf}
\end{eqnarray}
and 
\begin{equation}
Z_g Z_A^{1/2} Z_c =1 \;, \label{nrg}
\end{equation}
due to the non-renormalization theorem of the ghost-antighost-gluon vertex  in the Landau gauge \cite{Piguet:1995er}. Also, from the absence of one-loop counterterm of the kind $\theta^2 A^a_\mu U^a_\mu$, it follows that 
\begin{equation}
\theta^2_0 A^a_{0\mu} U^a_{0\mu} = \theta^2 A^a_\mu U^a_\mu \;. \label{thetaz}
\end{equation}
Therefore, from 
\begin{eqnarray}
\theta^2_0 & = & Z_{\theta^2} \theta^2 \;, \nonumber  \\ 
 A^a_{0\mu} & = & Z^{1/2}_A A^a_{\mu} \;, \nonumber \\
 U^a_{0\mu} & = & Z^{1/2}_A U^a_{\mu} \;, \label{rentheta}
\end{eqnarray}
we get 
\begin{equation}
Z_{\theta^2} =  Z_A^{-1} \;, \label{nrtheta} 
\end{equation} 
meaning that the renormalization factor\footnote{A purely algebraic proof, valid to all orders,  of the non-renormalization properties of the soft parameter $\theta^2$, eq.(\ref{nrtheta}), is under investigation \cite{wprogr}.} of the soft parameter $\theta^2$ can be expressed in terms of the gluon renormalization factor $Z_A$\footnote{We remark here that the non-renormalization of the soft coupling term, eq.\eqref{thetaz}, seems to be a general feature of theories leading to  Gribov type propagators, eqs.\eqref{paa}, \eqref{puu},\eqref{pau}. We observe in fact that a similar result holds in the Gribov-Zwanziger theory, for which the corresponding soft coupling term is given by
\begin{equation}
\gamma^2 g f^{abc} A^a_\mu \left( \phi^{bc}_{\mu}- {\bar \phi}^{bc}_\mu \right) \;, \label{gzsoft}
\end{equation}
where $\gamma$ is the Gribov parameter and $(\phi^{bc}_{\mu}, {\bar \phi}^{bc}_\mu)$ are auxiliary fields, see   \cite{Dudal:2009bf}. The renormalization factors of the Gribov parameter and of the auxiliary fields can be found in   
\cite{Dudal:2010fq}, and read
\begin{eqnarray}
\gamma^2_0 & = & Z_{\gamma^2} \gamma^2 \;, \nonumber \\
\phi^{bc}_{0\mu} & = & Z^{1/2}_\phi \phi^{bc}_\mu  \;,\nonumber \\
{\bar \phi}^{bc}_{0\mu} & = & Z^{1/2}_\phi {\bar \phi}^{bc}_\mu  \;, \label{garenf}
\end{eqnarray}
with 
\begin{equation}
Z_{\gamma^2} = Z^{-1/2}_g Z^{-1/4}_A \;, \qquad  Z^{1/2}_\phi= Z^{-1/2}_g Z^{-1/4}_A \;. \label{fgz}
\end{equation}
As a consequence
\begin{equation}
\gamma^2_0 g_0 f^{abc}A^a_{0\mu} \left( \phi^{bc}_{0\mu}- {\bar \phi}^{bc}_{0\mu} \right) = 
\gamma^2 g f^{abc}A^a_\mu \left( \phi^{bc}_{\mu}- {\bar \phi}^{bc}_\mu \right) \;, \label{gzsoftnr}
\end{equation}
which expresses the non-renormalization of the soft coupling term of the Gribov-Zwanziger action.}. The identity \eqref{p1gb} can be employed to generalize the one-loop argument to higher orders, enabling us to argue that the only divergences occurring in our model are those corresponding to $\theta^2=0$, {\it i.e.} those of the Yang-Mills theory in the Landau gauge.

\subsubsection{The Slavnov-Taylor identities} 

The existence of a soft breaking of the BRST symmetry does not prevent us to establish a set of Slavnov-Taylor identities which are suitable for an all-order algebraic analysis  of the renormalizability properties of the model \cite{wprogr}. The usual way of proceeding\footnote{See, for example, the case of the Slavnov-Taylor identities derived in the Gribov-Zwanziger theory  \cite{Dudal:2008sp,Dudal:2009bf}.} is that of  introducing an extended action which incorporates all local composite operators entering the soft breaking, by coupling them to a suitable set of external sources. The original action is thus recovered when the  sources acquire a particular value, which we shall refer to as the physical value. The renormalizability of the extended action entails thus the renormalizability of the starting action $S$, eq.\eqref{model}. \\\\In order to be able to discuss the non-renormalization property, eq.\eqref{nrtheta}, of the soft parameter $\theta^2$, we follow the procedure outlined in \cite{Dudal:2002pq} in the case of the study of the composite operator $A^a_\mu A^a_\mu$ in Yang-Mills theory, and introduce a set of sources $(J, \eta_\mu, \tau_\mu, N, M, \sigma^a_\mu, \rho^a_\mu)$ transforming as 
\begin{eqnarray} 
s J & = & 0 \;, \nonumber \\
s \tau_\mu & = & \eta_\mu  \;, \nonumber \\
s \eta_\mu & = & -\partial_\mu J  \;, \nonumber \\
s\rho^a_\mu & = & \sigma^a_\mu  \;, \nonumber \\
s \sigma^a_\mu & = & 0 \;, \nonumber \\
s M & = & N  \;, \nonumber \\
s N & = &J \;. \label{so1}
\end{eqnarray} 
The extended action $S^{ext}$ turns out to be
\begin{equation} 
S^{ext} = S\Bigl|_{\theta=0} + S_J \;, \label{ext}
\end{equation} 
where $S\Bigl|_{\theta=0}$ is defined in eq.\eqref{plm} and $S_J$ is given by 
\begin{eqnarray} 
S_J  & = & \int d^4x \left( J A^a_\mu U^a_\mu + \frac{\xi}{2} J^2 + \eta_\mu (c^a U^a_\mu + \omega^a A^a_\mu) 
+ \tau_\mu \left( s(c^a U^a_\mu + \omega^a A^a_\mu) \right)  \right)  \nonumber \\
& {\ }{\ }{\ }{\ }{\ }{\ } + & \int d^4x \left(  N \left( g f^{acb} A^c_\mu U^a_\mu c^b + g f^{acb} U^c_\mu A^a_\mu \omega^b \right)  + M \left( s \left( g f^{acb} A^c_\mu U^a_\mu c^b + g f^{acb} U^c_\mu A^a_\mu \omega^b\right) \right)    \right)    \nonumber \\
& {\ }{\ }{\ }{\ }{\ }{\ } + & \int d^4x \left(  \sigma^a_\mu \left( g f^{abc} c^b U^c_\mu  + g f^{abc} \omega^b A^c_\mu  \right)  + \rho^a_\mu  \left( s \left(  g f^{abc} c^b U^c_\mu  + g f^{abc} \omega^b A^c_\mu       \right) \right)    \right)    \;,  \label{sj}
\end{eqnarray} 
where $\xi$ is a dimensionless parameter needed to account for the ultraviolet divergences present in the correlation function $\langle \left(A^a_\mu(x) U^a_\mu(x) \right) \left(A^b_\nu(y) U^b_\nu(y)\right) \rangle$.  From expression \eqref{sj} one easily checks that the starting action $S$, eq.\eqref{model}, is recovered from the extended action $S^{ext}$ when the sources  $(J, \eta_\mu, \tau_\mu, N, M, \sigma^a_\mu, \rho^a_\mu)$ acquire the physical values 
\begin{equation} 
J\Big|_{phys}  =  i\sqrt{2} \theta^2 \;, \label{jpv}
\end{equation}
\begin{equation}
(\eta_\mu, \tau_\mu, N, M, \sigma^a_\mu, \rho^a_\mu)\Big|_{phys}  =  0 \;, \label{spv} 
\end{equation}
so that 
\begin{equation}
S^{ext}\Big|_{phys} = S \;. \label{rec}
\end{equation}
In order to write down the Slavnov-Taylor identities, a second set of external sources, $(\Omega^a_\mu, {\bar \Omega}^a_\mu, L^a, {\bar L}^a)$, properly  coupled to the non-linear BRST transformations of the fields $(A^a_\mu, U^a_\mu, c^a, \omega^a)$, eqs.\eqref{brst}, has to be introduced \cite{Piguet:1995er}, namely 
\begin{equation}
S_{\Omega} = \int d^4x \left( -\Omega^a_\mu D^{ab}_\mu(A) c^b - {\bar \Omega}^a_{\mu} D^{ab}_\mu(U)\omega^b 
+ \frac{g}{2} f^{abc} L^a c^b c^c  + \frac{g}{2} f^{abc} {\bar L}^a \omega^b \omega^c \right) \;. \label{sext}
\end{equation}
Therefore, it turns out that the action $\Sigma$ 
\begin{equation} 
\Sigma = S^{ext} + S_\Omega = S\Bigl|_{\theta=0} + S_J + S_\Omega \;, \label{sig}
\end{equation}
obeys the Slavnov-Taylor identities 
\begin{equation}
{\cal S}(\Sigma) = 0 \;, \label{sto}
\end{equation} 
where 
\begin{eqnarray}
{\cal S}(\Sigma)  & = & \int d^4x \; \left( \frac{\delta \Sigma}{\delta A^a_\mu} \frac{\delta \Sigma}{\delta \Omega^a_\mu}   + \frac{\delta \Sigma}{\delta U^a_\mu} \frac{\delta \Sigma}{\delta {\bar \Omega}^a_\mu} + 
\frac{\delta \Sigma}{\delta c^a} \frac{\delta \Sigma}{\delta L^a_\mu} + \frac{\delta \Sigma}{\delta \omega^a} \frac{\delta \Sigma}{\delta {\bar L}^a}  + b^a  \frac{\delta \Sigma}{\delta {\bar c}^a}    
+ {\bar b}^a  \frac{\delta \Sigma}{\delta {\bar \omega}^a}  -J c^a  \frac{\delta \Sigma}{\delta {\bar b}^a}   \right)   \nonumber \\
& + & \int d^4x\; \left(       -J \omega^a  \frac{\delta \Sigma}{\delta {b}^a}   
+ \sigma^a_\mu  \frac{\delta \Sigma}{\delta {\rho}^a_\mu} + \eta_\mu  \frac{\delta \Sigma}{\delta {\tau}_\mu} 
- (\partial_\mu J) \frac{\delta \Sigma}{\delta {\eta}_\mu}
+ N  \frac{\delta \Sigma}{\delta M}  + J \frac{\delta \Sigma}{\delta N} \right)  \;. \label{stf}
\end{eqnarray}
Moreover,  besides the Slavnov-Taylor identities \eqref{sto}, the extended action $\Sigma$ fulfills other Ward identities, which can be found in  Appendix  \ref{wid}.  These Ward identities turn out to be very helpful for a purely algebraic proof \cite{Piguet:1995er}  of the renormalizability of the model to all orders, a topic which we shall present in detail in a forthcoming work \cite{wprogr}.

\subsection{A few words on the field $U^a_\mu$\;: relationship with the $i$-particles}
In order to provide a better understanding of the introduction of the mirror field $U^a_\mu$, let us work out the relationship between our model, eq.\eqref{model}, and the $i$-particles \cite{Baulieu:2009ha}, {\it i.e.} the pair of unphysical modes with complex masses $\pm i \sqrt{2} \theta^2$ associated to a confining Gribov type propagator.  To this end, let us consider the quadratic part of the action $S$ containing the two gauge fields $(A^a_\mu, U^a_\mu)$, namely 
\begin{equation} 
S_{quad} = \int d^4x\; \left( \frac{1}{2} A^a_\mu (-\partial^2) A^a_\mu + \frac{1}{2} U^a_\mu (-\partial^2) U^a_\mu + 
i\sqrt{2} \theta^2 A^a_\mu U^a_\mu \right) \;, \label{qd} 
\end{equation}
where we have already taken into account the Landau gauge conditions, $\partial_\mu A^a_\mu=0$ and $\partial_\mu U^a_\mu=0$. Expression \eqref{qd} can be cast in diagonal form by introducing the two field combinations
\begin{eqnarray} 
\lambda^a_\mu & = & \frac{1}{\sqrt{2}} \left( A^a_\mu + U^a_\mu \right) \;,  \nonumber \\
\eta^a_\mu & = & \frac{1}{\sqrt{2}} \left( A^a_\mu - U^a_\mu \right) \;.  \label{ip}
\end{eqnarray}
Therefore 
\begin{equation}
S_{quad} = \int d^4x\; \left( \frac{1}{2} \lambda^a_\mu (-\partial^2+i\sqrt{2}\theta^2) \lambda^a_\mu + \frac{1}{2} \eta^a_\mu (-\partial^2-i\sqrt{2}\theta^2) \eta^a_\mu   \right) \;, \label{qd} 
\end{equation}
which describes in fact the propagation of two unphysical modes with complex masses $\pm i\sqrt{2}\theta^2$:
\begin{eqnarray} 
\langle \lambda^a_\mu(k) \lambda^b_\nu(-k) \rangle & = &  \frac{1}{2} \langle (A^a_\mu(k) + U^a_\mu(k) )
(A^b_\nu(-k) + U^b_\nu(-k)) \rangle =  \delta^{ab} \frac{1}{k^2+i\sqrt{2}\theta^2} \left( \delta_{\mu\nu} -\frac{k_\mu k_\nu}{k^2} \right) \;, \nonumber \\
\langle \eta^a_\mu(k) \eta^b_\nu(-k) \rangle & = &      \frac{1}{2} \langle (A^a_\mu(k) - U^a_\mu(k) )
(A^b_\nu(-k) - U^b_\nu(-k)) \rangle  =         \delta^{ab} \frac{1}{k^2-i\sqrt{2}\theta^2} \left( \delta_{\mu\nu} -\frac{k_\mu k_\nu}{k^2} \right) \;. \label{ipprop}
\end{eqnarray}
These are precisely the $i$-particles corresponding to the Gribov propagators in eqs.\eqref{paa},  \eqref{puu}, \eqref{pau}. We see thus that the action \eqref{model} has a direct interpretation in terms of $i$-particles. As pointed out in \cite{Baulieu:2009ha}, the advantage of introducing the fields $(\lambda^a_\mu, \eta^a_\mu)$ relies on the fact that they turn out to be  helpful in order to construct local composite operators whose one-loop correlation functions exhibit the  K\"all\'{e}n-Lehmann spectral representation. This feature stems for the observation that the basic momentum integral corresponding to the one-loop Feynman diagram containing  one propagator of the $\lambda$-type and one propagator of the $\eta$-type, {\it i.e.}
\begin{equation} 
{\cal I}(k^2) = \int \frac{d^4p}{(2\pi)^4} \; \frac{1}{\left( (k-p)^2+i\sqrt{2}\theta^2 \right) \left(  p^2 -i \sqrt{2}\theta^2 \right)}  \;, \label{bint}
\end{equation} 
exhibits a nice spectral representation, as shown in great detail in  \cite{Baulieu:2009ha}:
\begin{equation} 
{\cal I}(k^2) - {\cal I}(0) = \int_{2\sqrt{2}\theta^2}^{\infty} d\tau \rho(\tau) \left( \frac{1}{\tau+k^2} -\frac{1}{\tau} \right)  \;, \label{iint}
\end{equation}
where the spectral density 
\begin{equation} 
\rho(\tau) = \frac{1}{16\pi^2} \frac{\sqrt{\tau^2-8\theta^4}}{\tau}  \;, \label{spectr}
\end{equation}
is positive in the range of integration\footnote{The subtraction of the factor ${\cal I}(0)$ in eq.\eqref{iint} is needed to account for the divergent character of expression \eqref{bint} in four dimensions.}. This relevant property enables us to introduce sensible operators with good analyticity properties. As an example, let us quote the operator 
\begin{equation} 
{\cal O}_{\lambda\eta} = \left( \partial_\mu \lambda^a_\nu - \partial_\nu \lambda^a_\mu \right) \left( \partial_\mu \eta^a_\nu - \partial_\nu \eta^a_\mu \right)  \;, \label{iop}
\end{equation} 
also extensively investigated in \cite{Baulieu:2009ha}, where its two-point correlation function has been shown to be cast in the form of a spectral representation with positive spectral function\footnote{Also here, a suitable subtraction to get rid of ultraviolet divergences is needed, see \cite{Baulieu:2009ha} for details.}:
\begin{eqnarray} 
\langle {\cal O}_{\lambda\eta}(k) {\cal O}_{\lambda\eta}(-k) \rangle & = & \int_{2\sqrt{2}\theta^2}^{\infty} d\tau \frac{\rho(\tau)}{\tau+k^2}   \;, \nonumber \\
\rho(\tau) & =   & 12(N^2-1) \frac{ \sqrt{\tau^2-8\theta^4}\;(8\theta^4+\tau^2)}{32\pi^2\tau}           
 \;. \label{id}
\end{eqnarray} 
Looking at the expressions of the Gribov propagators in eqs.\eqref{paa},  \eqref{puu}, \eqref{pau}, we might argue that, in the nonperturbative confining infrared region, gluons get affected by nonperturbative long-range effects which alter the form of the propagator in such a way that they cannot correspond to excitations of the physical spectrum of the theory. Though, these long-range effects have no consequences on the deep ultraviolet perturbative region, where the usual interpretation of gluons in terms of partons is reliable. We can argue that, in our model, these long-range effects are precisely encoded in the mirror field $U^a_\mu$, which is only softly coupled to the gauge field $A^a_\mu$, through the mixed mass term $\theta^2 A^a_\mu U^a_\mu$. This term plays a very relevant role in the low energy region, where it gives rise to a confining Gribov type gluon propagator. Moreover, in the deep ultraviolet region such a soft coupling is negligible and the two gauge fields $(A^a_\mu, U^a_\mu)$ belong to two disjoint and non-interacting theories, according to eq.\eqref{plm}. In this sense, the introduction of the mirror field $U^a_\mu$ can be seen as a simple tool to implement a confining mechanism for the gluon field $A^a_\mu$. Finally, the introduction of the other mirror fields $({\bar b}^a, \omega^a, {\bar \omega}^a)$ stems from the need of preserving the renormalizability of the model.

\section{Example of a local composite BRST invariant  operator displaying the K\"all\'{e}n-Lehmann spectral representation} \label{decent}
Let us face now the construction of a local composite operator whose two-point correlation function exhibits the K\"all\'{e}n-Lehmann spectral representation at one-loop order, while being invariant under the BRST transformations. To construct such an operator, we observe that the BRST operator $s$ in eq.\eqref{brst} has non-vanishing cohomology classes \cite{Piguet:1995er}  which, in the sector of zero ghost number, can be identified with BRST invariant local polynomials built with the field strengths $F^a_{\mu\nu}(A), U^a_{\mu\nu}(U)$ and their covariant derivatives. Moreover, noticing that the quantity 
\begin{equation}
F^a_{\mu\nu}(x)U^a_{\mu\nu}(x) \;, \label{nbrstinv}
\end{equation} 
is not BRST invariant\footnote{An elementary calculation gives 
\begin{equation} 
s F^a_{\mu\nu}(x)U^a_{\mu\nu}(x) = - gf^{acb}F^c_{\mu\nu}U^a_{\mu\nu} c^b - gf^{abc} F^a_{\mu\nu}U^c_{\mu\nu} \omega^b \;. \label{calc}
\end{equation}
.}, it follows that the lowest dimensional  BRST invariant operators which might be introduced and which have  well defined transformation properties under the discrete mirror symmetry, eq.\eqref{mirror}, are given by the following two independent combinations
\begin{eqnarray}
{\cal O}^{+}_{AU}(x) & = & \frac{1}{2} \left( F^2(x) + U^2(x) \right) = \frac{1}{2}  \left(   F^a_{\mu\nu}(x)F^a_{\mu\nu}(x)+ U^a_{\mu\nu}(x)U^a_{\mu\nu}(x)  \right)  \;, \label{oplus} \\
{\cal O}^{-}_{AU}(x) & = & \frac{1}{2} \left( F^2(x) - U^2(x) \right) = \frac{1}{2}  \left(   F^a_{\mu\nu}(x)F^a_{\mu\nu}(x) - U^a_{\mu\nu}(x)U^a_{\mu\nu}(x)  \right)  \;. \label{ominus} 
\end{eqnarray}
As it is easily checked, these two operators have different transformation properties with respect to the mirror symmetry \eqref{mirror}, {\it i.e.}
\begin{eqnarray} 
{\cal O}^{+}_{AU}(x) & \rightarrow & + \; {\cal O}^{+}_{AU}(x)  \;, \nonumber \\
{\cal O}^{-}_{AU}(x) & \rightarrow & - \; {\cal O}^{-}_{AU}(x)  \;. \label{mop}
\end{eqnarray}
Although the operator ${\cal O}^{-}_{AU}(x)$ changes sign with respect to the mirror symmetry, its two-point correlation function $\langle {\cal O}^{-}_{AU}(k) {\cal O}^{-}_{AU}(-k) \rangle$ is, of course, invariant. \\\\For reasons which will become clear shortly, we shall disregard the operator ${\cal O}^{+}_{AU}(x)$, focusing our attention to ${\cal O}^{-}_{AU}(x)$. We also point out that the two operators ${\cal O}^{+}_{AU}(x)$ and  ${\cal O}^{-}_{AU}(x)$ cannot mix at quantum level, as they have different transformation properties under the mirror symmetry, eqs.\eqref{mop}. The operator ${\cal O}^{-}_{AU}(x)$ can only mix  with  composite operators which are odd under the mirror symmetry. \\\\The reason why we keep the operator  ${\cal O}^{-}_{AU}(x)$  relies on the important property
that, to the lowest order in the fields, it reduces precisely to the $i$-particles operator ${\cal O}_{\lambda\eta}$ of eq.\eqref{iop}. In fact 
\begin{equation} 
{\cal O}^{-}_{AU}(x)  =  \frac{1}{2} \left( F^a_{\mu\nu} + U^a_{\mu\nu} \right) \left( F^a_{\mu\nu} - U^a_{\mu\nu} \right)  =  \left( \partial_\mu \lambda^a_\nu - \partial_\nu \lambda^a_\mu \right) \left( \partial_\mu \eta^a_\nu - \partial_\nu \eta^a_\mu \right)  + {\ }{\ }{ higher {\ }order{\ } terms} \;. \label{iopeq}
\end{equation}
As such, we can  immediately state that the two-point correlation function $\langle {\cal O}^{-}_{AU}(k) {\cal O}^{-}_{AU}(-k) \rangle$ exhibits the  K\"all\'{e}n-Lehmann representation at one-loop order. In fact
\begin{equation}
\langle {\cal O}^{-}_{AU}(k) {\cal O}^{-}_{AU}(-k) \rangle \Big|_{one-loop} = \langle {\cal O}_{\lambda\eta}(k) {\cal O}_{\lambda\eta}(-k) \rangle  =  12(N^2-1) \int_{2\sqrt{2}\theta^2}^{\infty} d\tau  
{\frac{ \sqrt{\tau^2-8\theta^4}\;(8\theta^4+\tau^2)}{32\pi^2\tau} }\frac{1}{\tau+k^2}
\;. \label{klrep}
\end{equation}
Concerning the operator ${\cal O}^{+}_{AU}(x)$, it turns out that it does not display the same analyticity properties of  ${\cal O}^{-}_{AU}(x)$. This can be checked by expressing it in terms of the fields $(\lambda^a_\mu, \eta^a_\mu)$:
\begin{equation} 
{\cal O}^{+}_{AU}(x)  =  \frac{1}{2} \left( \left(\partial_\mu \lambda^a_\nu - \partial_\nu \lambda^a_\mu \right)^2 +  \left( \partial_\mu \eta^a_\nu - \partial_\nu \eta^a_\mu \right)^2 \right)  + {\ }{\ }{ higher {\ }order{\ } terms.} \;
\label{plus}
\end{equation}
From expression \eqref{plus} it follows that, at one-loop order, the correlation function $\langle {\cal O}^{+}_{AU}(x) {\cal O}^{+}_{AU}(y) \rangle$ receives contributions from  terms of the kind  $\langle \;\left(\partial_\mu \lambda^a_\nu - \partial_\nu \lambda^a_\mu \right)^2_x \left(\partial_\sigma \lambda^b_\rho - \partial_\rho \lambda^b_\sigma \right)^2_y \;\rangle$ which gives rise to a Feynman diagram with two internal lines of the $\lambda^a_\mu$-type. For the resulting momentum integral one gets
\begin{equation} 
 \int \frac{d^4p}{(2\pi)^4} \; \frac{p^2(p-k)^2+2(p^2-pk)^2}{\left( (k-p)^2+i\sqrt{2}\theta^2 \right) \left(  p^2 +i \sqrt{2}\theta^2 \right)}  \;, \label{plu-bint}
\end{equation}
which does have unphysical cuts along the imaginary axis \cite{Baulieu:2009ha}. This is a consequence of the fact that, unlike expression  ${\cal I}(k^2)$ of eq.\eqref{bint}, the sign of the complex masses in the denominators of eq.\eqref{plu-bint} is the same. \\\\In summary, we have been able to introduce a BRST invariant local composite operator ${\cal O}^{-}_{AU}(x)$, given by expression \eqref{ominus}, whose two point correlation function exhibits nice spectral properties. Other examples of higher dimensional BRST invariant composite operators displaying good analyticity properties can be easily constructed by inserting an appropriate number of covariant derivatives into expression \eqref{ominus}, as:  
\begin{equation}
{\cal O}^{-}_{DFDU}= \frac{1}{2}\left(   \left(D^{ab}_\sigma(A)F^b_{\mu\nu}\right)\left(D^{am}_\sigma(A)F^m_{\mu\nu}\right) - \left(D^{ab}_\sigma(U)U^b_{\mu\nu}\right)\left(D^{am}_\sigma(U)U^m_{\mu\nu}\right)  \right)  \;. \label{o1} 
\end{equation}

\subsection{A short survey on the renormalizability of the operator ${\cal O}^{-}_{AU}$ } 

It remains the hard task of investigating the correlation function $\langle {\cal O}^{-}_{AU}(x) {\cal O}^{-}_{AU}(y) \rangle$ at higher orders, and check out if the analyticity properties found at one-loop still hold. Although this analysis is beyond the aim of the present work, let us provide here a few preliminary remarks about the renormalization of the composite operator ${\cal O}^{-}_{AU}$. \\\\The property of being BRST invariant will play an important role at the quantum level. Due to the fact that the BRST symmetry is softly broken, one expects that the operator ${\cal O}^{-}_{AU}$ will mix with two classes of composite operators, as also observed in \cite{Dudal:2009zh} in the case of the study of the renormalization of the composite operator $F^2(x)$ within the Gribov-Zwanziger theory. \\\\The first class is given by operators which have the same dimension of  ${\cal O}^{-}_{AU}$, {\it i.e.} dimension four. These operators are BRST invariant, as they are independent from the soft breaking parameter $\theta^2$. Taking into account that ${\cal O}^{-}_{AU}$ is odd under the discrete mirror symmetry, eq.\eqref{mop}, and that  ${\cal O}^{+}_{AU}$ and ${\cal O}^{-}_{AU}$ are the only two possible non-trivial elements of the cohomology of the BRST operator in the space of the local polynomials with dimension four and zero ghost number, it follows that the remaining invariant operators with which ${\cal O}^{-}_{AU}$ can mix are BRST exact, {\it i.e.} they can be written as the $s$-variation of suitable local field polynomials which are odd under the mirror symmetry\footnote{For completeness, we mention that composite operators proportional to the equations of motion might also show up. However, they do not contribute to the correlation functions \cite{Dudal:2009zh,Dudal:2008tg}. }. \\\\The second class of composite operators is given by operators of lower dimension, {\it i.e.} of dimension two, which depend explicitly on the soft  parameter $\theta^2$. These operators cannot  correspond to non-trivial BRST invariant elements, as there are no local non-trivial gauge invariant quantities of  dimension two. Taking into account that ${\cal O}^{-}_{AU}$ is odd under the mirror symmetry and that the ghosts and antighosts fields $(c^a, \omega^a, {\bar c}^a, {\bar \omega}^a)$ cannot enter due to the ghost and antighost Ward identities, eqs.\eqref{ghwid}, \eqref{antigh}, it follows that the only dimension two operator which can potentially show up is 
\begin{equation} 
{\cal O}^{-}_{2-dim}(x) = \frac{1}{2} \left(  A^a_\mu(x) A^a_\mu(x) - U^a_\mu(x) U^a_\mu(x) \right)  \;. \label{2dim}
\end{equation}  
It is interesting to observe that this operator has a nice interpretation in terms of $i$-particles, namely 
\begin{equation} 
{\cal O}^{-}_{2-dim}(x) =    \lambda^a_\mu(x) \eta^a_\mu(x)   \;. \label{2ip}
\end{equation} 
As such, the correlation function 
\begin{equation} 
\langle  {\cal O}^{-}_{2-dim}(x) {\cal O}^{-}_{2-dim}(y)  \rangle   =  \langle   (\lambda^a_\mu \eta^a_\mu)(x) \;      (\lambda^b_\nu \eta^b_\nu)(y)  \rangle  \;,  \label{n2d}
\end{equation}
will give rise to a momentum integral of the kind of ${\cal I}(k^2)$ in eq.\eqref{bint}. It will thus display the  K\"all\'{e}n-Lehmann spectral representation. The same feature holds for the correlation function describing the mixing between the two operators ${\cal O}^{-}_{AU}(x)$ and ${\cal O}^{-}_{2-dim}(x)$. In fact, to the first order, 
\begin{equation} 
\langle {\cal O}^{-}_{AU}(x) {\cal O}^{-}_{2-dim}(y) \rangle = \langle \left(\left( \partial_\mu \lambda^a_\nu - \partial_\nu \lambda^a_\mu \right) \left( \partial_\mu \eta^a_\nu - \partial_\nu \eta^a_\mu \right)\right)_x \left( \lambda^b_\sigma \eta^b_\sigma\right)_y \rangle   + {\ }{\ }{ higher {\ }order{\ } terms} \;. \label{nice}
\end{equation}
The Feynman diagram corresponding to expression \eqref{nice} will contain one internal line of the $\lambda^a_\mu$-type and one internal line of the $\eta^a_\mu$-type, giving rise to a momentum integral with a cut structure similar to that of ${\cal I}(k^2)$, eq.\eqref{bint}. These properties can be taken as an encouraging evidence that a good analytic structure might survive at higher orders.   \\\\Finally, let us observe that the operator ${\cal O}^{-}_{AU}(x)$ can be easily introduced  into the extended action $\Sigma$, eq.\eqref{sig}, through a BRST invariant source $q(x)$, {\it i.e.}
\begin{equation} 
\Sigma^{q} = \Sigma + \int d^4x\; q(x) {\cal O}^{-}_{AU}(x) \;, \label{q-act}
\end{equation}
where the source $q(x)$ is odd under the mirror symmetry, $q(x) \rightarrow -q(x)$. The action $\Sigma^q$ is seen to obey the Slavnov-Taylor identities \eqref{sto} 
\begin{equation}
{\cal S}(\Sigma^q) = 0 \;, \label{st}
\end{equation} 
as well as all other Ward identities given in Appendix \ref{wid}, allowing thus for an all order algebraic analysis of the renormalizability of ${\cal O}^{-}_{AU}(x)$ along the lines outlined in  \cite{Dudal:2009zh,Dudal:2008tg}.

\section{A possible way to take into account the Gribov copies. The scaling type solution for the gluon and ghost propagators} \label{Gribov}

The action $S$ of our model, eq.\eqref{model}, contains a mass parameter $\theta^2$ which, till now, has been considered as a free parameter. Willing to give a better physical interpretation of the model, a dynamical framework for the parameter $\theta^2$ should be provided, {\it i.e.} we should be able to establish a gap equation enabling us to express $\theta^2$ as a function of the coupling constant $g$, as in the case of the Gribov parameter $\gamma$ of the Gribov-Zwanziger theory  \cite{Gribov:1977wm,Zwanziger:1988jt,Zwanziger:1989mf,Zwanziger:1992qr}. \\\\A possible way to achieve a meaningful gap equation for $\theta^2$ relies on the observation that the gauge fixed action $S$ is plagued by the existence of the Gribov copies \cite{Gribov:1977wm}, as it is apparent from expression \eqref{model} where the Landau gauge fixing conditions, $\partial_\mu A^a_\mu=0$, $\partial_\mu U^a_\mu=0$, have been employed. \\\\In \cite{Gribov:1977wm}, it has been  suggested that, in order to get rid of the Gribov copies, the domain of integration in the functional integral should be restricted to the so-called Gribov region $\Omega_A$, which is defined as the set of field configurations which fulfill the Landau gauge condition and for which the Faddeev-Popov operator, ${\cal M}^{ab}(A)=-\partial_\mu \left( \delta^{ab} \partial_\mu + g f^{acb} A^{c}_\mu \right)$, is strictly positive, namely 
\begin{equation}
 \Omega_A = \Big\{A^a_\mu; \; \partial_\mu A^a_\mu=0\;,\;{\cal M}^{ab}(A)=-\partial_\mu \left( \delta^{ab} \partial_\mu + g f^{acb} A^{c}_\mu \right) >0 \;\Big\} \;. \label{gbregion}
 \end{equation}  
The region $\Omega_A$ has been proven to be convex and bounded in all directions in field space \cite{Dell'Antonio:1989jn}. Moreover, every gauge orbit passes through $\Omega_A$ \cite{Dell'Antonio:1991xt}\footnote{Nowadays, it is known that the Gribov region $\Omega_A$ is not completely free from Gribov copies \cite{vanBaal:1991zw}. Additional equivalent gauge field configurations exist within $\Omega_A$. To get rid of these additional copies, a further restriction of the domain of integration to a smaller region, known as the fundamental modular region,  should be implemented. This region is contained within the Gribov region $\Omega_A$ and is known to be free from Gribov copies. Nevertheless, so far, a way to implement the restriction of the domain of integration in the functional integral to this region in a local and renormalizable way has not yet been achieved. Here,  we limit ourselves to the restriction to the Gribov region $\Omega_A$.}. Its boundary, $\partial \Omega_A$, where the first vanishing eigenvalue of the Faddeev-Popov operator shows up, is known as the Gribov horizon. \\\\In our case, due to the mirror symmetry \eqref{mirror}, there is also a mirror Gribov region $\Omega_U$:
\begin{equation}
 \Omega_U = \Big\{U^a_\mu; \; \partial_\mu U^a_\mu=0\;,\;{\cal M}^{ab}(U)=-\partial_\mu \left( \delta^{ab} \partial_\mu + g f^{acb} U^{c}_\mu \right) >0 \;\Big\} \;. \label{mgregion}
 \end{equation} 
Evidently, $\Omega_U$ enjoys the same properties of $\Omega_A$. \\\\In his original work \cite{Gribov:1977wm}, Gribov was able to implement the restriction to the region $\Omega_A$ by demanding that the parameter $\gamma$ fulfills  a particular gap equation, known as the no-pole condition, which follows as a consequence of the proper definition of the region $\Omega_A$. Here,  we shall follow the same route, {\it i.e.} we shall employ Gribov's no-pole condition in order to establish the gap equation for the parameter $\theta^2$. \\\\Let us proceed by giving a short account on Gribov's no-pole condition, see also  \cite{Sobreiro:2005ec}  for a pedagogical introduction. It relies on the observation that, within the regions $\Omega_A, \Omega_U$, the Faddeev-Popov operators are invertible and their inverse, $ {\cal M}^{-1}(A)$ and $ {\cal M}^{-1}(U)$, are positive definite, as follows from the definition of the Gribov regions, eqs.\eqref{gbregion},\eqref{mgregion}.  Keeping in mind that $ {\cal M}^{-1}(A) $  and $ {\cal M}^{-1}(U)$  are nothing but the two-point ghost functions, {\it i.e.} $ {{\cal M}_A}^{-1}(x,y) = {\cal G}_{{\bar c}c}(x,y) = \langle {\bar c}(x) c(y) \rangle $, $ {{\cal M}_U}^{-1}(x,y) = {\cal G}_{{\bar \omega}\omega}(x,y) = \langle {\bar \omega}(x) \omega(y) \rangle $, we infer that, within the regions $\Omega_A$ and $\Omega_U$,  ${\cal G}_{{\bar c}c}(x,y)$  and ${\cal G}_{{\bar \omega}\omega}(x,y)$ stay always positive. \\\\According to  \cite{Gribov:1977wm}, we now parametrize the  two point ghost functions ${\cal G}_{{\bar c}c}(k)$,  ${\cal G}_{{\bar \omega}\omega}(k)$ in momentum space as 
\begin{equation} 
{\cal G}^{ab}_{{\bar c}c}(k) = {\cal G}^{ab}_{{\bar \omega} \omega}(k) = \frac{\delta^{ab}}{k^2} \frac{1}{1-\sigma(k^2,\theta^2)} \;, \label{ghf}
\end{equation}
where the equality  ${\cal G}^{ab}_{{\bar c}c}(k) = {\cal G}^{ab}_{{\bar \omega} \omega}(k)$ is a consequence of the mirror symmetry \eqref{mirror}. Gribov's no-pole condition amounts to require that the form factor $\sigma(k^2,\theta^2)$ is bounded by one, $\sigma(k^2,\theta^2) \le 1$, so that expression \eqref{ghf} cannot have a pole for a non-vanishing value of the momentum $k$  \cite{Gribov:1977wm}. Expression \eqref{ghf} stays thus always positive, namely the Gribov horizon $\partial \Omega_A$ is never crossed. The only allowed singularity is at $k=0$, whose meaning is that of approaching the horizon $\partial \Omega_A$, where ${\cal G}^{ab}_{{\bar c}c}(k)$ is singular due to the appearance of zero modes of the Faddeev-Popov operator $ {\cal M}(A) $. Of course, the same features hold for the mirror two point ghost function ${\cal G}^{ab}_{{\bar \omega}\omega}(k)$. \\\\Following \cite{Gribov:1977wm}, the no-pole condition is implemented by stating that
\begin{equation}
\sigma(0,\theta^2)=1 \;, \label{nopole}
\end{equation}
 which yields the gap equation determining the parameter $\theta^2$. To the first order, the form factor $\sigma(k^2,\theta^2)$ is easily evaluated and found to be \cite{Gribov:1977wm,Dudal:2008sp} 
 \begin{eqnarray}
 \sigma(k^2,\theta^2) & = & \frac{N}{N^2-1} \frac{g^2}{k^2} \int \frac{d^dq}{(2\pi)^d} \frac{(k-q)_\mu k_\nu}{(k-q)^2} \langle A^a_\mu(q) A^a_\nu(-q) \rangle  \nonumber \\
 & = & Ng^2 \frac{k_\mu k_\nu}{k^2} \int \frac{d^dq}{(2\pi)^d} \frac{q^2}{(k-q)^2(q^4+2\theta^4)}\left( \delta_{\mu\nu} -\frac{q_\mu q_\nu}{q^2} \right) \;, \label{ffact}
 \end{eqnarray}
 where dimensional regularization, $d=4-\varepsilon$, has been employed. Therefore, at one-loop order, the gap equation \eqref{nopole} yields
 \begin{equation}
 1 = Ng^2\; \frac{d-1}{d} \int \frac{d^dq}{(2\pi)^d} \frac{1}{q^4+2\theta^4} \;, \label{gapp}
 \end{equation}
which enables us to express the parameter $\theta^2$ as a function of the coupling constant $g$. 
\\\\Equation \eqref{gapp} also provides us a better understanding of the meaning of the mass paramater $\theta^2$. We notice that, in practice, Gribov's no-pole condition amounts to impose a boundary condition on the two-point ghost function ${\cal G}^{ab}_{{\bar c}c}(k)$, {\it i.e.} one requires that ${\cal G}^{ab}_{{\bar c}c}(k)$ stays always positive, so that the Gribov horizon is never crossed. In its original work \cite{Gribov:1977wm}, this boundary condition was implemented by introducing a certain massive parameter $\gamma^2$, known as the Gribov parameter, whose value was fixed by the requirement of positivity of the ghost propagator, resulting in a gap equation which has precisely the same form of the gap equation \eqref{gapp} obeyed by our massive parameter $\theta^2$, see, for instance, eq.(43) of \cite{Gribov:1977wm}. This is a relevant feature of our model, meaning that no new mass parameters are needed to ensure positivity of the ghost propagator, and thus the restriction to the Gribov region $ \Omega_A$. This is achieved by requiring that the parameter $\theta^2$ acquires a particular value, determined precisely by the gap equation  \eqref{gapp}. In other words, a unique mass parameter is sufficient in order to ensure positivity of the ghost propagator. As such, our massive parameter $\theta^2$ plays  the same role of the parameter $\gamma^2$ of Gribov's original work, a fact which is also apparent by remarking that both parameters give rise to the same gluon propagator, namely 
\begin{eqnarray} 
\langle A^a_\mu(k) A^b_\nu(-k) \rangle_{Gribov} & = & \delta^{ab} \frac{k^2}{k^4+\gamma^4} \left( \delta_{\mu\nu} -\frac{k_\mu k_\nu}{k^2} \right) \;, \nonumber \\
\langle A^a_\mu(k) A^b_\nu(-k) \rangle_{replica} & = & \delta^{ab} \frac{k^2}{k^4+2\theta^4} \left( \delta_{\mu\nu} -\frac{k_\mu k_\nu}{k^2} \right) \;. \label{paaboth}
\end{eqnarray}
From the gap equation  \eqref{gapp} it follows that, for small values of the momentum $k$, $k \approx  0$, 
\begin{equation} 
\sigma(k^2,\theta^2) \Big|_{k^2 \approx 0} = 1 - {\cal C} k^2 \;, \label{kzero}
\end{equation}
for some constant factor $\cal C$\footnote{Equation \eqref{kzero} can be derived by making use of eq.\eqref{gapp}, written in the form 
\begin{equation}
Ng^2\;  \int \frac{d^dq}{(2\pi)^d} \frac{1}{q^4+2\theta^4}\left( \delta_{\mu\nu} -\frac{q_\mu q_\nu}{q^2} \right) = \delta_{\mu\nu}  \;.  \label{gapf1}
\end{equation}
Therefore, for the form factor $\sigma(k^2,\theta^2)$ in eq.\eqref{ffact}, we get
\begin{eqnarray}
\sigma(k^2,\theta^2) & = & 1 + Ng^2 \frac{k_\mu k_\nu}{k^2}  \int \frac{d^dq}{(2\pi)^d} \left( \delta_{\mu\nu} -\frac{q_\mu q_\nu}{q^2} \right) \left(  \frac{q^2}{(k-q)^2} -1   \right) \frac{1}{q^4+2\theta^4}  \nonumber \\
& = &  1 - Ng^2 \frac{k_\mu k_\nu}{k^2}  \int \frac{d^dq}{(2\pi)^d} \left( \delta_{\mu\nu} -\frac{q_\mu q_\nu}{q^2} \right) \frac{1}{q^4+2\theta^4}  \frac{k^2-2kq}{(k-q)^2}  \;, \label{insert}
\end{eqnarray}
from which eq.\eqref{kzero} follows.} Thus, for the behavior of the ghost propagators ${\cal G}^{ab}_{{\bar c}c}(k)$, ${\cal G}^{ab}_{{\bar \omega}\omega}(k)$ near the origin in momentum space, we get 
\begin{equation}
{\cal G}^{ab}_{{\bar c}c}(k)\Big|_{k\approx 0} = {\cal G}^{ab}_{{\bar \omega}\omega}(k)\Big|_{k\approx 0} \approx \frac{1}{k^4}  \;, \label{enh}
\end{equation}
 showing that imposing the no-pole condition  \eqref{gapp} yields  ghost propagators which are enhanced in the infrared. We have thus recovered the so-called scaling solution, {\it i.e.} a suppressed gluon propagator which vanishes at the origin, eq.\eqref{paa}, and enhanced ghosts, eq.\eqref{enh}.  This is precisely the type of solution which emerges from the Gribov-Zwanziger theory \cite{Gribov:1977wm,Zwanziger:1988jt,Zwanziger:1989mf,Zwanziger:1992qr,Gracey:2006dr,Zwanziger:2010iz}. A scaling type solution is also found in the studies of the Schwinger-Dyson equations  \cite{von Smekal:1997is,von Smekal:1997vx,Alkofer:2000wg}, provided the ghost enhancement is required by imposing a suitable boundary condition, see \cite{Fischer:2008uz,Huber:2009tx}  for a recent discussion of this topic. 
 
\section{Clarifying the difference between the Gribov-Zwanziger action and the replica model}\label{clar}
\subsection{The Gribov-Zwanziger action}
It is worth adding here a few additional remarks in order  to clarify in a  better way the difference between the Gribov-Zwanziger construction and the present  model.  \\\\As it has been already underlined, within the Gribov-Zwanziger approach, a nonperturbative formulation of Yang-Mills theory is obtained by restricting the domain of integration in the path integral to the Gribov region $\Omega_A$, as expressed by the partition function
\begin{equation} 
{\cal Z}_{GZ} = \int_{\Omega_A} [DA] \; \delta(\partial A^a) \; det(\partial_\mu D^{ab}_\mu \delta^4(x-y)) \; e^{-\frac{1}{4} \int d^4x\; F^a_{\mu\nu} F^a_{\mu\nu} }  \;. \label{gzpf}
\end{equation}
As shown in \cite{Gribov:1977wm,Zwanziger:1988jt,Zwanziger:1989mf}, the restriction to the region $\Omega_A$ is equivalent to the introduction of the horizon function  $S_h$, namely 
\begin{equation}
{\cal Z}_{GZ} = \int  [DA] \; \delta(\partial A^a) \; det(\partial_\mu D^{ab}_\mu \delta^4(x-y)) \; e^{-\left( \frac{1}{4} \int d^4x\; F^a_{\mu\nu} F^a_{\mu\nu}   + S_h \right)}\;. \label{gzpf}
\end{equation}  
where the horizon term\footnote{Although the expression of  $S_h$, eq.\eqref{hzz}, is non-local, we  remind that it can be cast in local form by means of the introduction of a suitable set of localizing fields. The resulting local action can be proven to be renormalizable to all orders \cite{Dudal:2009bf}. }  $S_h$ is given by 
 \begin{equation} 
 S_h =\gamma^4 \int d^4x \;h(x) =  \gamma^4 g^2 \int d^4x f^{amc} A^m_\mu \left( (-\partial_\mu D_\mu)^{-1} \right)^{ad} f^{dpc}A^p_\mu \label{hzz}
 \end{equation} 
 and $\gamma^2$ is the Gribov parameter, which is determined by the so-called no-pole condition \cite{Gribov:1977wm}. This condition, when expressed in terms of the horizon function $S_h$, takes the form \cite{Zwanziger:1988jt,Zwanziger:1989mf}
 \begin{equation} 
 \langle h(x) \rangle = 4 (N^2-1)  \;. \label{hz}
 \end{equation}
 At  one-loop order, the gap equation \eqref{hz} reads
 \begin{equation} 
 1 = g^2 N \frac{d-1}{d} \int \frac{d^dk}{(2\pi)^d} \frac{1}{k^4+2g^2N\gamma^4} \;. \label{hzone} 
 \end{equation} 
Taken together, equations \eqref{gzpf}, \eqref{hzz}, \eqref{hz} constitute the so called Gribov-Zwanziger framework for the nonperturbative formulation of Yang-Mills theory which takes into account the Gribov issue.   \\\\It is important to emphasize here that the gap equation \eqref{hz} is part of the nonperturbative definition of the Faddeev-Popov quantization formula within the Gribov-Zwanziger approach. In other words, the parameter $\gamma^2$ is not a free parameter of the theory. It is generated in a dynamical way, and its expression in terms of the gauge coupling constant $g$ is uniquely determined by \eqref{hz}. In particular, one has no more the freedom of setting $\gamma=0$, as one already sees, for example, from the one-loop approximation \eqref{hzone}. Equations \eqref{gzpf}, \eqref{hzz} provide a nonperturbative definition of the quantized Yang-Mills theory only when the parameter $\gamma^2$ is constrained by the gap equation \eqref{hz}. This has a very transparent physical meaning. In fact, when $\gamma^2$ obeys the gap equation \eqref{hz} it cannot be treated as a free parameter. Instead, it becomes a function of the gauge coupling constant $g$ and of $\Lambda_{QCD}$ \cite{Gribov:1977wm,Zwanziger:1988jt,Zwanziger:1989mf}. As such, expression  \eqref{gzpf} has exactly the same number of parameters of Yang-Mills theory. This ensures that the restriction of the domain of integration in the path integral to the Gribov region $\Omega_A$ does not bring us outside of Yang-Mills theory.    
\subsection{The nonperturbative definition of the replica model} 
At this stage, the difference between the Gribov-Zwanziger construction and our approach should be manifest. We point out that the nonperturbative definition of the replica model is encoded in the following 
 three equations, \eqref{zrep}, \eqref{repact}, \eqref{np11}:  
\begin{equation}
{\cal Z}_{replica} = \int [DA Db Dc D{\bar c}] \; [DU D{\bar b} D\omega D{\bar \omega}] \; e^{-S}  \; \label{zrep}
\end{equation} 
with $S$ being the action of the replica model 
 \begin{eqnarray}
S  & = &  \int d^4x\; \left(  \frac{1}{4} F^a_{\mu\nu}(A) F^a_{\mu\nu}(A) + \frac{1}{4} U^a_{\mu\nu}(U) U^a_{\mu\nu}(U)
     + i \sqrt{2} \theta^2 A^a_\mu U^a_\mu \right) \nonumber \\
    & + &  \int d^4x \; \left(b^a \partial_\mu A^a_\mu + {\bar c}^a \partial_{\mu} D^{ab}_\mu(A) c^b
    +  {\bar b}^a \partial_\mu U^a_\mu + {\bar \omega}^a \partial_{\mu} D^{ab}_\mu(U) \omega^b \right)  \;. \label{repact}
\end{eqnarray}
 and $\theta^2$ obeying the gap equation \eqref{nopole}, namely 
\begin{equation}
\sigma(0,\theta^2)=1 \;.  \label{np11}
\end{equation}
Taken together, expressions \eqref{zrep}, \eqref{repact}, \eqref{np11} constitute our definition of the nonperturbative version of the replica model. In particular, as it happens in the case of the Gribov parameter $\gamma^2$, also the parameter $\theta^2$ is now completely fixed in terms of the gauge coupling constant $g$ through the gap equation  \eqref{np11}. In particular, after imposing the gap equation \eqref{nopole}, one has no more the freedom to set  $\theta^2=0$, as it is already apparent from the one-loop approximation \eqref{gapp}. \\\\Of course, we are not claiming here that the two partition functions ${\cal Z}_{GZ}$ and ${\cal Z}_{replica}$ will give rise to the same physics.  We remind here that the main motivation to study the replica model is that of investigating the consequences of the soft breaking of the BRST symmetry in a confining theory exhibiting a gluon propagator of the Gribov type. The replica model has to be seen as a useful simpler model allowing us to investigate in an analytic way aspects of the gluon confinement. \\\\ We do argue, however,  that, taken together,  the three expressions \eqref{zrep}, \eqref{repact}, \eqref{np11} provide a nonperturbative definition of the replica model which takes into proper account the issue of the Gribov copies, as implied by the positivity of the corresponding ghost propagators. As such, the partition function ${\cal Z}_{replica}$ does not need to be further restricted to the Gribov region $\Omega_A$, {\it i.e.} there is no need for the Gribov-Zwanziger construction to be implemented again.  This would lead to a complete different model which turns out to be plagued by various difficulties, as it will be illustrated in the next section.  
\subsection{Combining the replica model and the Gribov-Zwanziger action} 
One might  argue that, somehow, the replica model and the Gribov-Zwanziger action could be combined into a single theory. This would amount to start with two decoupled Yang-Mills theories, {\it i.e.} with  $\theta^2=0$, and implement the restriction to the Gribov region $\Omega_A$ by means of the Gribov-Zwanziger construction. At the end, the parameter $\theta^2$ is re-introduced in order to couple in a soft way the two gauge fields $A^a_\mu$ and $U^a_\mu$. The final action compatible with the mirror symmetry contains  two massive parameters $\gamma^2, \theta^2$, and reads
 \begin{eqnarray}
S_{replica-GZ}  & = &  \int d^4x\; \left(  \frac{1}{4} F^a_{\mu\nu}(A) F^a_{\mu\nu}(A) + \frac{1}{4} U^a_{\mu\nu}(U) U^a_{\mu\nu}(U) + S_h(A) + S_h(U) 
     + i \sqrt{2} \theta^2 A^a_\mu U^a_\mu \right) \nonumber \\
    & + &  \int d^4x \; \left(b^a \partial_\mu A^a_\mu + {\bar c}^a \partial_{\mu} D^{ab}_\mu(A) c^b
    +  {\bar b}^a \partial_\mu U^a_\mu + {\bar \omega}^a \partial_{\mu} D^{ab}_\mu(U) \omega^b \right)  \;. \label{repGZ}
\end{eqnarray}        
with $S_h(A)$ given in eq.\eqref{hzz}. Although being a confining model, to our understanding, the action \eqref{repGZ} would represent a completely different model, leading to several difficulties of not easy solution. The first one is encoded in the form of the gluon propagator, which takes now a rather complicated form 
\begin{equation}
\langle A^a_\mu(k) A^b_\nu(-k) \rangle = \delta^{ab} \left( \delta_{\mu\nu} - \frac{k_\mu k_\nu}{k^2} \right) \frac{ k^2(k^4+\gamma^4) }{k^8 + 2k^4(\gamma^4+\theta^4) + \gamma^8}  \;. \label{newprop}
\end{equation} 
This propagator is not of the Gribov type and does not exhibit a simple  $i$-particles decomposition. As a consequence, the explicit analytic computation of the spectral densities of the BRST invariant operators discussed in the previous sections looks prohibitive. A second difficulty arises at the level of the gap equations which should be imposed in order to obtain the values of the two parameters $\gamma^2, \theta^2$. Even if the no-pole condition 
\begin{equation}
\sigma(0,\theta^2,\gamma^2)=1 \;, \label{nopole-rgz}
\end{equation}
is required in order to ensure positivity of the ghost propagators, it should be noticed that it will be not sufficient to determine both parameters  $\gamma^2, \theta^2$. A second gap equation would be required. This is not an easy matter. Imposing a gap equation is a highly nontrivial issue, which has to be properly motivated by physical and/or geometrical  considerations, as in the case of the no-pole condition \eqref{nopole-rgz} which follows from the necessity of handling the Gribov copies.  At present, it does not look easy to propose a  second gap equation in order to determine both parameters uniquely.

\section{A refined version of the model. The decoupling type solution} \label{refined}
As already mentioned, the most recent lattice numerical simulations \cite{Cucchieri:2007rg,Cucchieri:2008fc,Cucchieri:2008mv,Cucchieri:2009zt,Bogolubsky:2009dc,Bogolubsky:2009qb,Dudal:2010tf} point towards a gluon propagator which is suppressed in the infrared and which attains a finite non-vanishing value at zero momentum, while the ghost propagator turns out to be not enhanced, keeping an almost free behavior, ${\cal G}^{ab}_{{\bar c}c}(k) \Big|_{k^2\approx 0}\approx \frac{1}{k^2} $.  \\\\This behavior, known as the decoupling solution, has also been reported from the analysis of the Schwinger-Dyson equations \cite{Aguilar:2004sw,Aguilar:2008xm,Boucaud:2008ky}. It can be seen as arising from a different choice of the boundary condition for the ghost propagator \cite{Fischer:2008uz}. \\\\Within the context of the Gribov-Zwanziger theory, the decoupling solution has been proven to show up \cite{Dudal:2007cw,Dudal:2008sp,Dudal:2008rm}  when the non-perturbative effects associated to the dynamics of the additional fields $(\phi^{ab}_{\mu}, {\bar \phi}^{ab}_\mu, \omega^{ab}_{\mu}, {\bar \omega}^{ab}_{\mu} )$ needed to implement the restriction to the Gribov region in a local and renormalizable way are taken into account. The resulting action is known as the refined Gribov-Zwanziger action  \cite{Dudal:2007cw,Dudal:2008sp,Dudal:2008rm}.  \\\\In this section, we intend to show that  a refined version of our model can also be introduced. It gives rise to a gluon propagator which does not vanish at zero momentum,  while preserving locality and renormalizability. This feature relies on the possibility of modifying the starting action $S$, eq.\eqref{model}, by adding a kind of Curci-Ferrari mass term  \cite{Curci:1976bt}, {\it i.e.} a mass term whose BRST variation vanishes on-shell, see  \cite{Delduc:1989uc,Dudal:2003pe,Wschebor:2007vh}. In our case, taking into account the mirror symmetry \eqref{mirror}, this mass term reads 
\begin{equation}
S_{m} = \frac{m^2}{2}  \int d^4x\; \left( A^a_\mu A^a_\mu + U^a_\mu U^a_\mu \right) \;. \label{cfmass}
\end{equation}
Accordingly, the action of the refined version of the model is
\begin{eqnarray}
S_{ref} & = & S_{FP} + S_{MFP} + S_{\theta}  + S_m   = S + S_m  \nonumber \\
    & = &  \int d^4x\; \left(  \frac{1}{4} F^a_{\mu\nu}(A) F^a_{\mu\nu}(A) + \frac{1}{4} U^a_{\mu\nu}(U) U^a_{\mu\nu}(U)
     + i \sqrt{2} \theta^2 A^a_\mu U^a_\mu  + 
 \frac{m^2}{2}  \left( A^a_\mu A^a_\mu + U^a_\mu U^a_\mu \right)   
     \right) \nonumber \\
    & + &  \int d^4x \; \left(b^a \partial_\mu A^a_\mu + {\bar c}^a \partial_{\mu} D^{ab}_\mu(A) c^b
    +  {\bar b}^a \partial_\mu U^a_\mu + {\bar \omega}^a \partial_{\mu} D^{ab}_\mu(U) \omega^b \right)  \;. \label{refmodel}
\end{eqnarray}
As it is easily checked, the BRST variation of the mass term $S_m$, eq.\eqref{cfmass}, turns out to be proportional to the equations of motion. In fact 
\begin{equation} 
s S_m = m^2 \int d^4x \left(  -A^a_\mu \partial_\mu c^a - U^a_\mu \partial_\mu \omega^a  \right)  = m^2 
\int d^4x \left( c^a \frac{\delta S_{ref}}{\delta b^a} + \omega^a \frac{\delta S_{ref}}{\delta {\bar b}^a}  \right) \;, \label{onshell}
\end{equation}
so that it vanishes on-shell. This property gives rise to a modified version of the Slavnov-Taylor identities which account for the renormalizability of the refined action $S_{ref}$ \cite{wprogr}. \\\\For the gluon propagator, we have now 
\begin{equation}
\langle A^a_\mu(k) A^b_\nu(-k) \rangle = \delta^{ab} \frac{k^2+m^2}{(k^2+m^2)^2+2\theta^4} \left( \delta_{\mu\nu} -\frac{k_\mu k_\nu}{k^2} \right) \;, \label{refpaa}
\end{equation}
which is of the type already found in the refined Gribov-Zwanziger action \cite{Dudal:2007cw,Dudal:2008sp,Dudal:2008rm}\footnote{We mention here that a slightly more general expression for the gluon propagator can be obtained from the generalized mass term 
\begin{equation}
S_{m\mu} =  \int d^4x\; \left( \frac{m^2}{2}A^a_\mu A^a_\mu + \frac{\mu^2}{2} U^a_\mu U^a_\mu \right) \;, \label{gcfmass}
\end{equation}
which also enjoys the property of being BRST invariant on-shell. For the gluon propagator, we obtain 
\begin{equation}
\langle A^a_\mu(k) A^b_\nu(-k) \rangle = \delta^{ab} \frac{k^2+\mu^2}{k^4 + (m^2+\mu^2)k^2 +(2\theta^4+ m^2\mu^2) } \left( \delta_{\mu\nu} -\frac{k_\mu k_\nu}{k^2} \right) \;. \label{grefpaa}
\end{equation} }. Clearly, expression \eqref{refpaa} attains a finite nonvanishing value at zero momentum. \\\\Even if the issues of a possible dynamical origin of the second mass parameter $m^2$ in eq.\eqref{cfmass} and of its consequences on the infrared behavior of the ghost propagators are not addressed in the present work, the possibility that the decoupling type behavior for the gluon propagator, eq.\eqref{refpaa}, can be accommodated for represents an interesting feature of our model. It is worth observing that, as in the case of the Gribov propagator \eqref{gbp}, the refined propagator \eqref{refpaa} also allows for the introduction of $i$-particles, namely
\begin{equation} 
\frac{1}{(k^2+m^2)^2+2\theta^4}  = \frac{1}{(k^2+m^2)+i\sqrt{2} \theta^2} \frac{1}{(k^2+m^2)-i\sqrt{2} \theta^2} \;. 
\label{mipart}
\end{equation}
In particular, the analysis of the analyticity properties of the BRST invariant operator ${\cal O}^{-}_{AU}(x)$, eq.\eqref{ominus}, generalizes to the case of the refined action \eqref{refmodel}. 

\section{Conclusion} \label{conclusion}

In this work a few aspects related to gluon confinement have been addressed within the Euclidean quantum field theory. \\\\We have presented a gauge model, eq.\eqref{model}, whose main feature is that of exhibiting a soft breaking of the BRST symmetry, encoded in the mass parameter $\theta^2$, eq.\eqref{mass}.  This soft breaking  is responsible for the confining behavior of the propagators of the elementary fields, eqs.\eqref{paa}, \eqref{puu}, \eqref{pau}. Furthermore, we have been able to show that examples of BRST invariant composite operators whose one-loop correlation functions exhibit the K{\"a}ll{\'e}n-Lehamnn spectral representation can be introduced.  \\\\It is worth to spend a few words on the strategy which has been used in the construction of our model. We point out that the mechanism of the soft breaking of the BRST symmetry has played a double role. On one side, it is responsible for the confinement of the elementary fields, as expressed by the Gribov propagators, eqs.\eqref{paa}, \eqref{puu}, \eqref{pau}. On the other side, it has enabled us to introduce examples of composite operators whose one-loop correlation functions turn out to be compatible with the unitarity requirement. After that, the existence of the Gribov phenomenon, which is an intrinsic feature of the gauge fixing procedure, has been employed to derive the gap equation for the soft parameter associated to the breaking of the BRST symmetry. \\\\To some extent, this framework provides a different way of looking at the Gribov issue. In other words, one first pay attention to the construction of a confining model compatible with the requirements of renormalizability and of unitarity, encoded in the existence of a set of renormalizable composite operators with good analyticity properties. As we have seen, the resulting model turns out to depend on the soft parameter responsible for the breaking of the BRST symmetry. As such, it is spoiled of any physical interpretation, unless one is able to derive a suitable gap equation allowing for a dynamical determination of the aforementioned soft parameter. This is precisely done by invoking the Gribov phenomenon, {\it i.e.} the existence of the Gribov copies is used as a powerful tool in order to achieve the gap equation for the soft parameter. Said otherwise, after one is able to construct a renormalizable and unitary confining gauge theory, the Gribov phenomenon is taken as a physical input in order to provide a dynamical framework for the soft parameter responsible for the breaking of the BRST symmetry, thus making possible the contact with $QCD$. Within this context, it would be interesting to pursue the investigation on the analyticity properties displayed by our model, by checking out if these properties can be extended to higher orders.  \\\\The main conclusion which emerges from the present work is that the mechanism of the soft breaking of the BRST symmetry might be helpful in order to account for several aspects of a confining theory. Certainly, much work is needed in order to achieve a satisfactory understanding of the issue of the BRST symmetry versus the Gribov horizon. Nevertheless, the example of the gauge model which we have presented here suggests that a possible reconciliation between the BRST symmetry and the presence of the Gribov horizon through the mechanism of the soft breaking might provide a useful and interesting pathway to gluon confinement.

\section*{Acknowledgments}
It is a pleasure to thank Laurent Baulieu, David Dudal, Marcelo Guimaraes, Markus Huber, Nele Vandersickel and Daniel Zwanziger for stimulating conversations. \\\\The Conselho Nacional de Desenvolvimento Cient\'{\i}fico e
Tecnol\'{o}gico (CNPq-Brazil), the Faperj, Funda{\c{c}}{\~{a}}o de
Amparo {\`{a}} Pesquisa do Estado do Rio de Janeiro, the Latin
American Center for Physics (CLAF), the SR2-UERJ,  the
Coordena{\c{c}}{\~{a}}o de Aperfei{\c{c}}oamento de Pessoal de
N{\'{\i}}vel Superior (CAPES) and the Espa{\c c}o X, are gratefully acknowledged.

  \newpage

 \begin{appendix}
\section{Ward Identities} \label{wid}
As we have already mentioned in Subsection \ref{ren}, the action $\Sigma$, eq.\eqref{sig}, fulfills a rather large set of Ward identities, which we shall enlist below. Let us first give the full expression of $\Sigma$, namely 
\begin{eqnarray} 
\Sigma & = &  \int d^4x \left(  \frac{1}{4} F^a_{\mu\nu}(A) F^a_{\mu\nu}(A) + \frac{1}{4} U^a_{\mu\nu}(U) U^a_{\mu\nu}(U) + b^a \partial_\mu A^a_\mu + {\bar c}^a \partial_{\mu} D^{ab}_\mu(A) c^b +  {\bar b}^a \partial_\mu U^a_\mu + {\bar \omega}^a \partial_{\mu} D^{ab}_\mu(U) \omega^b \right) \nonumber \\
    & + &  \int d^4x  \left( J A^a_\mu U^a_\mu +\frac{\xi}{2} J^2 +\eta_\mu \left(c^a U^a_\mu + \omega^a A^a_\mu \right) + \tau_\mu \left(   \frac{g}{2} f^{acb}\left( U^a_\mu c^c c^b +A^a_\mu \omega^c \omega^b\right) + c^a D^{ab}_\mu(U)\omega^b  + \omega^a D^{ab}_\mu(A)c^b\right)  \right)  \nonumber \\ 
& + &  \int d^4x  \left(  gf^{acb} N \left(  A^c_\mu U^a_\mu c^b + U^c_\mu A^a_\mu \omega^b  \right)   
- gf^{acb} M \left( (\partial_\mu c^c) U^a_\mu c^b + A^c_\mu (\partial_\mu \omega^a)c^b 
+ (\partial_\mu \omega^c) A^a_\mu \omega^b + U^c_\mu (\partial_\mu c^a)\omega^b      \right)     \right)  \nonumber \\ 
& + &  \int d^4x  \left(  gf^{abc} \sigma^a_\mu \left( c^b U^c_\mu +\omega^b A^c_\mu \right)    
+ gf^{abc} \rho^a_\mu \left(   \frac{gf^{bmn}}{2} \left(  c^m c^n U^c_\mu + \omega^m \omega^n A^c_\mu \right)  
+ c^b D^{cm}_\mu(U)\omega^m + \omega^b D^{cm}_\mu(A)c^m  \right)  \right)  \nonumber \\
& + &  \int d^4x  \left(  -\Omega^a_\mu D^{ab}_\mu(A) c^b  -{\bar \Omega}^a_\mu D^{ab}_\mu(U) \omega^b 
+\frac{g}{2} f^{abc} L^a c^b c^c + \frac{g}{2}f^{abc} {\bar L}^a \omega^b \omega^c   \right)     \;. 
\label{act}
\end{eqnarray}
It turns out that $\Sigma$ obeys the following Ward identities: 
\begin{itemize} 
\item the Slavnov-Taylor identities, eq.\eqref{sto}, 
\begin{equation}
{\cal S}(\Sigma) = 0 \;, \label{st1}
\end{equation} 
\begin{eqnarray}
{\cal S}(\Sigma)  & = & \int d^4x \; \left( \frac{\delta \Sigma}{\delta A^a_\mu} \frac{\delta \Sigma}{\delta \Omega^a_\mu}   + \frac{\delta \Sigma}{\delta U^a_\mu} \frac{\delta \Sigma}{\delta {\bar \Omega}^a_\mu} + 
\frac{\delta \Sigma}{\delta c^a} \frac{\delta \Sigma}{\delta L^a_\mu} + \frac{\delta \Sigma}{\delta \omega^a} \frac{\delta \Sigma}{\delta {\bar L}^a}  + b^a  \frac{\delta \Sigma}{\delta {\bar c}^a}    
+ {\bar b}^a  \frac{\delta \Sigma}{\delta {\bar \omega}^a}  -J c^a  \frac{\delta \Sigma}{\delta {\bar b}^a}   \right)   \nonumber \\
& + & \int d^4x\; \left(       -J \omega^a  \frac{\delta \Sigma}{\delta {b}^a}   
+ \sigma^a_\mu  \frac{\delta \Sigma}{\delta {\rho}^a_\mu} + \eta_\mu  \frac{\delta \Sigma}{\delta {\tau}_\mu} 
- (\partial_\mu J) \frac{\delta \Sigma}{\delta {\eta}_\mu}
+ N  \frac{\delta \Sigma}{\delta M}  + J \frac{\delta \Sigma}{\delta N} \right)  \;. \label{stf1}
\end{eqnarray}
\item the ghost Ward identity \cite{Blasi:1990xz,Piguet:1995er}
\begin{equation} 
{\cal G}^m \Sigma = \Delta^m_{cl}  \;,  \label{ghwid}
\end{equation}
where 
\begin{eqnarray}
{\cal G}^m & = & \int d^4x \left( \frac{\delta}{\delta c^m} + \frac{\delta}{\delta \omega^m} + gf^{acm}{\bar c}^a \frac{\delta}{\delta b^c} + gf^{acm} {\bar \omega}^a \frac{\delta}{\delta {\bar b}^c}  
+ \tau_\mu \left(  \frac{\delta}{\delta \Omega^m} + \frac{\delta}{\delta {\bar \Omega}^m}   \right)      \right)  \nonumber \\
& + & \int d^4x \left(    gf^{acm} \rho^a_\mu \frac{\delta}{\delta \sigma^c_\mu}  
- gf^{acm} \rho^a_\mu \left(  \frac{\delta}{\delta \Omega^c} + \frac{\delta}{\delta {\bar \Omega}^c}   \right)   \right)  \;, \label{ghop}
\end{eqnarray}
and 
\begin{equation}
\Delta^m_{cl} = \int d^4x \left( gf^{acm} \left(  \Omega^a_\mu A^c_\mu  + {\bar \Omega}^a_\mu U^c_\mu 
-L^a c^c - {\bar L}^a \omega^c + \sigma^a_\mu U^c_\mu + \sigma^a_\mu A^c_\mu\right)   
- \eta_\mu \left( U^m_\mu + A^m_\mu \right)   \right)  \;. \label{ghb}
\end{equation}
Notice that expression \eqref{ghb} is linear in the quantum fields. As such, $\Delta^m_{cl}$ is a classical breaking, not affected by quantum corrections \cite{Piguet:1995er}.
\item the gauge conditions \cite{Piguet:1995er}
\begin{equation}
\frac{\delta \Sigma}{\delta b^a} = \partial_\mu A^a_\mu \;, \qquad  \frac{\delta \Sigma}{\delta {\bar b}^a} = \partial_\mu U^a_\mu  \;, \label{gcond}
\end{equation}
\item the antighost equations  \cite{Piguet:1995er}
\begin{equation} 
\frac{\delta \Sigma}{\delta {\bar c}^a} + \partial_\mu \frac{\delta \Sigma}{\delta {\Omega}^a_\mu} =0  \;, 
\qquad 
\frac{\delta \Sigma}{\delta {\bar \omega}^a} + \partial_\mu \frac{\delta \Sigma}{\delta {\bar \Omega}^a_\mu} =0  \;. \label{antigh} 
\end{equation}
\item the discrete mirror symmetry 
\begin{eqnarray}
A^a_{\mu}  \rightarrow   U^a_{\mu} \;,  & \qquad & 
U^a_{\mu}  \rightarrow A^a_{\mu} \;, \nonumber \\
b^a   \rightarrow   {\bar b}^a \;, & \qquad &  
{\bar b}^a  \rightarrow { b}^a \;, \nonumber \\
c^a  \rightarrow {\omega}^a \;,&  \qquad &  
\omega^a  \rightarrow {c}^a \;, \nonumber \\
{\bar c}^a  \rightarrow {\bar \omega}^a \;, & \qquad &  
{\bar \omega}^a  \rightarrow {\bar c }^a \;, \nonumber \\
\Omega^a_\mu \rightarrow {\bar \Omega}^a_\mu \;, & \qquad & {\bar \Omega}^a_\mu \rightarrow { \Omega}^a_\mu
\;, \nonumber \\
L^a  \rightarrow {\bar L}^a \;, & \qquad & {\bar L}^a \rightarrow { L}^a 
\;, \nonumber \\
J \rightarrow J \;,  & \qquad & \eta_\mu \rightarrow \eta_\mu \;,  \nonumber \\
\tau_\mu \rightarrow \tau_\mu \;,  & \qquad & N  \rightarrow N \;, \nonumber \\
\sigma^a_\mu \rightarrow \sigma^a_\mu \;,  & \qquad & \rho^a_\mu   \rightarrow \rho^a_\mu  \;, \nonumber \\
M \rightarrow M \;.  \nonumber \\
\label{mirror1}
\end{eqnarray}

\end{itemize}   
These Ward identities constitute a powerful set up for a purely algebraic investigation \cite{Piguet:1995er} of the renormalizability properties of the model to all orders \cite{wprogr}.

\end{appendix}

\end{document}